\documentclass[lettersize,journal]{IEEEtran}

\usepackage{cite}
\usepackage{amsmath,amssymb,amsfonts}
\usepackage{graphicx}
\usepackage{algorithm}
\usepackage{algorithmic}
\usepackage{caption}
\usepackage{subcaption}
\usepackage{multirow}
\usepackage{pifont}
\usepackage{booktabs}
\usepackage[all]{nowidow}
\usepackage{fancyhdr}
\usepackage{xcolor}
\usepackage[hidelinks]{hyperref}

\begin{document}
\bstctlcite{IEEEexample:BSTcontrol}

\title{FORTALESA: Fault-Tolerant Reconfigurable Systolic Array for DNN Inference}

\author{
    \IEEEauthorblockN{
    Natalia Cherezova\IEEEauthorrefmark{1},
    Artur Jutman\IEEEauthorrefmark{1}\IEEEauthorrefmark{2},
    Maksim Jenihhin\IEEEauthorrefmark{1}\\
    }
    \IEEEauthorblockA{\IEEEauthorrefmark{1}Department of Computer Systems, Tallinn University of Technology, Tallinn, Estonia}\\
    \IEEEauthorblockA{\IEEEauthorrefmark{2}Testonica Lab, Tallinn, Estonia}\\
    \IEEEauthorblockA{\{natalia.cherezova, maksim.jenihhin\}@taltech.ee, artur@testonica.com}
}

\pagestyle{fancy}
\fancyhf{}
\fancyhead[L]{\footnotesize\copyright 2025. This manuscript version is made available under the CC-BY-NC-ND 4.0 license https://creativecommons.org/licenses/by-nc-nd/4.0/}

\maketitle
\thispagestyle{fancy}
\fancyhf{}
\fancyhead[L]{\footnotesize\copyright 2025. This manuscript version is made available under the CC-BY-NC-ND 4.0 license https://creativecommons.org/licenses/by-nc-nd/4.0/}

\begin{abstract}
The emergence of Deep Neural Networks (DNNs) in mission- and safety-critical applications brings their reliability to the front. High performance demands of DNNs require the use of specialized hardware accelerators. Systolic array architecture is widely used in DNN accelerators due to its parallelism and regular structure. This work presents a run-time reconfigurable systolic array architecture with three execution modes and four implementation options. All four implementations are evaluated in terms of resource utilization, throughput, and fault tolerance improvement. The proposed architecture is used for reliability enhancement of DNN inference on systolic array through heterogeneous mapping of different network layers to different execution modes. The approach is supported by a novel reliability assessment method based on fault propagation analysis. It is used for the exploration of the appropriate execution mode--layer mapping for DNN inference. The proposed architecture efficiently protects registers and MAC units of systolic array PEs from transient and permanent faults. The reconfigurability feature enables a speedup of up to $3\times$, depending on layer vulnerability. Furthermore, it requires $6\times$ fewer resources compared to static redundancy and $2.5\times$ fewer resources compared to the previously proposed solution for transient faults.
\end{abstract}

\begin{IEEEkeywords}
systolic array, fault tolerance, deep neural networks, accelerator.
\end{IEEEkeywords}


\section{Introduction}
Recent years have been marked by a tremendous increase in adoption of Machine Learning (ML) algorithms, specifically Deep Neural Networks (DNNs), and their employment in mission- and safety-critical application domains, such as automotive, avionics, and space. The computational complexity of DNN models requires the use of specialized hardware accelerators. Many DNN accelerators are based on systolic array architecture, including Google TPU \cite{TPU}, MIT Eyeriss \cite{Eyeriss}, and Berkeley Gemmini \cite{Gemmini}. However, as nanoelectronics used in hardware accelerators are prone to faults caused by radiation and aging, the concern about their reliability comes to the front \cite{Ahmadilivani2024}.

Furthermore, with the current need for self-aware computing systems that can monitor their state and adapt to changing environments and operating conditions, an important property of hardware design is the ability to adjust or reconfigure to changing conditions for improved availability. To address those needs, we propose a run-time reconfigurable systolic array architecture that tackles the balance between reliability and performance through different execution modes. As a target application, we consider the inference of DNNs. Since different DNN components have different vulnerability levels, they might benefit from heterogeneous mapping to different execution modes of the proposed systolic array. We also introduce a novel reliability assessment method for systolic arrays. It is used to determine the appropriate execution mode--layer mapping and consequently to validate the effectiveness of the proposed architecture.

The contributions of the paper are
\begin{itemize}
    \item a systolic array architecture with three execution modes (FORTALESA) for run-time reconfigurable redundancy and flexible reliability vs. performance trade-off;
    \item hardware implementation and evaluation of the proposed architecture;
    \item a novel reliability assessment method for DNN inference on a systolic array based on fault propagation analysis;
    \item a flexible DNN protection methodology balancing reliability and performance when executed on a systolic array.
\end{itemize}

For DNN protection, different network layers are mapped to appropriate execution modes of the systolic array based on their vulnerability. The proposed reliability assessment method is used for the exploration of the execution mode--layer mapping.

The proposed architecture efficiently protects both sequential (registers) and combinational components of systolic array processing elements. While the main objective is to address transient faults (soft errors), the proposed architecture can handle permanent faults as well, allowing for graceful degradation. The reconfigurability feature of FORTALESA enables a speedup of up to $3\times$, depending on layer vulnerability, and requires $6\times$ fewer resources compared to static redundancy and $2.5\times$ fewer resources than the previously proposed solution for transient faults.

The rest of the paper is organized the following way. Section~\ref{related_works} gives an overview of related works. Section~\ref{background} presents the background on systolic array architecture. Section~\ref{architecture} introduces the architecture of the proposed reconfigurable systolic array. Section~\ref{reliability_assessment} explains the proposed reliability assessment method for systolic arrays based on fault propagation. Section~\ref{evaluation} evaluates the effectiveness of the proposed methodology for DNN reliability, and Section~\ref{conclusion} concludes the paper.

\section{Related works}\label{related_works}
A reconfigurable redundancy approach is common in multicore processors, e.g., ARM's Triple Core Lock-Step (TCLS) system \cite{Iturbe2018} and RISC-V based multicore processing cluster with on-demand redundancy for efficient performance vs. reliability trade-off \cite{Rogenmoser2022}. Such solutions allow the use of individual cores separately to run different tasks for high performance or as a group to run the same task for high reliability. Execution mode can be reconfigured during run-time, thus allowing the computing system to adapt to changing conditions and task requirements.

While there are several ways to ensure the reliability of DNN inference, the most common approaches imply redundancy, fault-aware training, and algorithm-based fault tolerance (ABFT). Redundancy indicates replication of the components working in parallel (spatial redundancy) or re-execution of the same operation (temporal redundancy). Since redundancy incurs high overhead, the goal is to find a trade-off between reliability and performance by identifying and protecting more vulnerable components. To tackle this challenge, prior works have proposed selective replication of the layers \cite{Libano2019, Bolchini2022}, of the channels within a layer \cite{Bertoa2023}, of the weights \cite{Khoshavi2020} or neurons \cite{Ahmadilivani2023}. Another common approach requires re-training of the DNN with the added faults to improve the resilience of the network. Fault-aware training was used to address faults in FPGAs \cite{Zahid2020}, timing \cite{Jaicnao2015} and computational \cite{Xu2019} errors in accelerators. ABFT methods consider certain features of the application for fault detection and correction. Proposed solutions include checksums \cite{Ozen2020a, Ozen2020b}, bounded activation functions \cite{Hoang2020}, restricted output ranges \cite{Chen2021} and quantile shifts \cite{Geissler2023}, and AN codes \cite{Goldstein2021}.

Systolic arrays were first introduced in the 1980s \cite{Kung1982} and recently became popular for accelerating DNNs. As the reliability of DNN accelerators comes to the front, recent works have proposed methods for the assessment and enhancement of systolic arrays. \cite{Vacca2023} analyzed the effects of transient faults in the combinational logic of systolic arrays. \cite{Tan2023} presented Saca-FI, a microarchitecture-level fault injection framework for systolic array-based CNN accelerators, for the analysis of transient faults in registers. Saca-FI is based on SCALE-Sim, a cycle-accurate simulator for DNN inference on systolic arrays \cite{Samajdar2020}. \cite{Agarwal2023} proposed an RTL-level fault injection framework for the analysis of stuck-at faults within multiply and accumulate (MAC) units. \cite{Taheri2024} presented a fault injection framework that models systolic array dataflow using Uniform Recurrent Equations. Unlike previous works, the proposed reliability assessment method is based on fault propagation analysis. Instead of injecting faults in the microarchitectural or RTL model of a systolic array, which is a very time-consuming task; errors are added directly to the DNN layer output using fault propagation analysis that allows to significantly speed up the fault injection process.

Reliability enhancement methods for systolic arrays, similarly to DNNs, include ABFT and network retraining. \cite{Zhao2022} presented a permanent fault correction technique by introducing a re-computing unit that recalculates the computations assigned to faulty processing elements. \cite{Lee2024b} addressed permanent faults through bypassing faulty units and assigning their computations to the redundant ones. Similarly, \cite{Lee2023} used pruning of faulty processing elements for permanent fault correction. \cite{Zhang2019} proposed a fault-aware pruning plus retraining (FAP+T) technique to address manufacturing defects in the systolic array-based accelerators. It requires modification of the target DNN architecture through pruning and retraining to adapt it to a specific defect in the chip. \cite{Libano2023} presented an ABFT technique for permanent fault detection in systolic arrays on FPGAs. Permanent faults in configuration memory are considered in that work. Another permanent fault detection approach assuming faults in registers and MAC units is presented in \cite{Vacca2023a}. The approach is based on checksum calculations and was implemented on FPGA as well. \cite{Cora2025} presented a fault tolerance solution for an FPGA-based systolic array accelerator. Faults in configuration memory affecting the systolic array datapath are detected using checksum values and corrected using dynamic partial reconfiguration of the FPGA. \cite{Peltekis2024, Peltekis2025} presented an ABFT methodology specifically developed for sparse systolic arrays, considering their unique characteristics. While most works address permanent faults, \cite{Safarpour2022} proposed an ABFT method for detecting timing faults due to reduced voltage. Several works proposed test architectures for permanent and transition-delay fault detection in low-power accelerators based on systolic arrays \cite{Ibtesam2022, Lee2025}. Compared to the aforementioned works that mostly address detection of permanent faults, the proposed reconfigurable systolic array architecture addresses both permanent and transient faults, enables reliability-performance trade-off, provides fault correction in both fault-tolerant execution modes, and does not require retraining of a DNN for each particular chip.

Reconfigurable DNN accelerators based on a systolic array have been proposed before: to improve energy efficiency \cite{Yin2018}, to decrease inference latency \cite{Lee2024} and power consumption \cite{Peltekis2023}, to boost transformer models performance \cite{Zhao2023}. In contrast to the previous works, the proposed reconfigurable systolic array architecture improves reliability of the DNN inference while balancing performance.

\section{Background}\label{background}
\subsection{Systolic array architecture}
Systolic array has a regular structure consisting of an array of Processing Elements (PEs). Each PE consists of a Multiply-Accumulate (MAC) unit and registers to hold the intermediate values. Memory buffers for activations and weights are located at the top and left sides of the systolic array respectively. Commonly, weights are fed vertically from top to bottom, thus each column is assigned to one output channel, and activations (inputs) are fed from left to right, hence each row processes values from the same sliding windows. Dataflow type defines the internal structure of PEs and the movement of data through the array. Dataflow types include weight-stationary (WS), input-stationary (IS), and output-stationary (OS) \cite{Chen2016}.

In the weight-stationary dataflow, the weight matrix is first pre-loaded into a systolic array by loading one row per cycle, then the input matrix propagates through the array from left to right. The outputs (results of MAC operation) propagate from top to bottom. In this scenario, weights are stationary and are kept in the systolic array during matrix multiplication execution. This dataflow maximizes the reuse of the weights.

In the input-stationary dataflow, activations are pre-loaded in the systolic array similarly to WS and kept in the systolic array during the matrix multiplication process. The data movement and structure of the systolic array are similar to WS dataflow.

In the output-stationary dataflow, outputs are kept in the PEs of the systolic array, while weights and activations are streaming through the systolic array. Weights are streaming from top to bottom, and activations (inputs) are streaming from left to right. Outputs are accumulated inside PEs. This dataflow type does not need a pre-loading stage.

An \textbf{output-stationary systolic array} is considered for this work. It is commonly used and allows high inputs and outputs reuse \cite{Xu2021, Deng2020} and, on average, has higher performance \cite{Lee2024}. Previous works have shown that it is more fault-tolerant \cite{Burel2022}.

\subsection{Convolution on systolic array}
The convolution operation is mapped to the systolic array by transforming it to matrix multiplication using im2col algorithm \cite{Chellapilla2006}. An input tensor is transformed into a $(H_{out} \cdot W_{out}) \times (H_k \cdot W_k \cdot C_{out})$ matrix and weights are transformed into a $(H_k \cdot W_k \cdot C_{in}) \times C_{out}$ matrix, where $H_{out}$ and $W_{out}$ are the height and width of the output tensor, $H_k$ and $W_k$ are the height and width of a kernel, $C_{in}$ and $C_{out}$ are the number of channels in the input and output tensor correspondingly. When inputs and weights are represented this way, each column of the output-stationary systolic array calculates values for one output channel, and each row calculates values of the same pixel across all output channels.

If the input matrices exceed the size of the systolic array, matrix multiplication is performed in tiles corresponding to the size of the systolic array grid.

The latency $L$ of a matrix multiplication on a systolic array is calculated as
\begin{equation}
    L_{SA} = M + N - 1 + N - 1 = M + 2N - 2\,,
\end{equation}
where $N \times N$ is the size of the systolic array and $N \times M$ is the size of the matrices \cite{Samajdar2020}.

The number of tiles in each matrix is calculated as
\begin{align}
    T_a &= \left\lceil \frac{P}{N} \right\rceil \,,\\
    T_w &= \left\lceil \frac{K}{N} \right\rceil \,,
\end{align}
where $T_a$ and $T_w$ are the number of tiles in activations and weights correspondingly, $P$ is the size of the reshaped activations, and $K$ is the number of output channels and the size of the weights matrix. The total number of steps needed to calculate the output equals $S = T_a \cdot T_w$, and the total latency of matrix multiplication on the systolic array is therefore
\begin{equation}
    L_{total} = S \cdot L_{SA} = \left\lceil \frac{P}{N} \right\rceil \cdot \left\lceil \frac{K}{N} \right\rceil \cdot (M + 2N - 2).
\end{equation}

\section{Fault-tolerant reconfigurable systolic array}\label{architecture}
We propose FORTALESA, a systolic array with reconfigurable redundancy that provides flexible reliability--performance trade-off for critical tasks requiring fault tolerance and non-critical ones that benefit from increased parallelism.

The proposed architecture supports three \textbf{execution modes}:
\begin{itemize}
    \item no redundancy for increased performance;
    \item dual-redundancy grouping (DRG) for reliability--performance trade-off;
    \item triple-redundancy grouping (TRG) for full protection.
\end{itemize}
The selected execution mode is configured through a control signal that controls multiplexers. The control signal comes from the host processor based on the task requirement. Since different layers of DNNs have different vulnerability, it is possible to execute different layers in different modes, thus achieving an efficient trade-off between performance and reliability.

The proposed architecture addresses faults in registers and arithmetic units of systolic array PEs, since they lead to silent data corruption without indication of an error, unlike faults in control logic that usually result in functional interrupts that signal that there was a fault and require a reset.

\begin{figure}[h]
    \centering
    \includegraphics[width=0.95\linewidth]{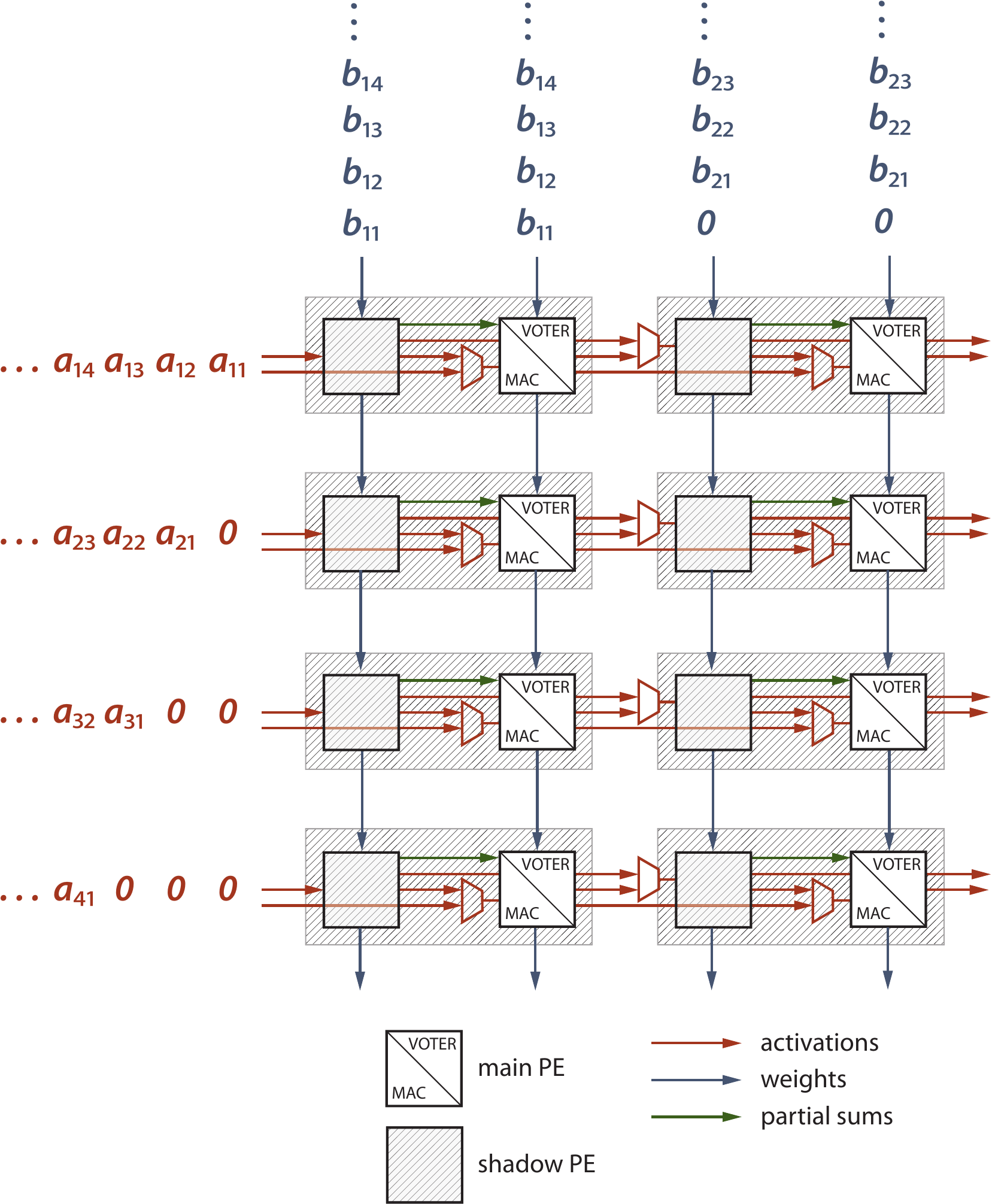}
    \caption{Arrangement of the systolic array in the dual-redundancy grouping execution mode (DRGA or DRG0)}
    \label{fig:dmr_mode}
\end{figure}

\textbf{DRG execution mode.} In the DRG execution mode, two neighboring PEs form a group (Fig.~\ref{fig:dmr_mode}). They receive the same inputs, one of them acts as the main PE, the other as a shadow. Main PE compares two partial sums (OREG) and adjusts its partial sum value without re-execution using one of the possible fault correction techniques: (a) averaging of two values (DRGA) or (b) setting faulty bits to zero (DRG0). While dual redundancy does not enable full protection, these correction techniques decrease the effect of the fault on the output values, thus enabling a reliability--performance trade-off. The choice of the fault correction technique is a design-time parameter that defines two implementation options for the DRG execution mode.

Additional connections are introduced to support redundant execution modes. To ensure that faulty IREG or WREG values do not propagate through the array, each PE passes the activation to the next PE of the same type, skipping one in between. While faults in buffers and interconnects are not in the focus of this work, such interleaved connections can help protect against those faults as well. During the fault-tolerant execution modes, there are two copies of input data traversing through the systolic array. The shadow PE also passes the value of the OREG (partial sum) to the main PE for comparison. Main PE has additional functionality for fault correction that works in parallel with the MAC unit. Since the updated partial sum is available on the next clock cycle, the latency of the systolic array working in DRG execution mode equals the latency of the $N \times N/2$ systolic array plus 1:
\begin{equation}
    L_{SA}^{drg} = M + N - 1 + \frac{N}{2} - 1 + 1 = M + \frac{3N}{2} - 1\,.
\end{equation}

The total latency is then
\begin{equation}
    L_{total}^{drg} = \left\lceil \frac{P}{N} \right\rceil \cdot \left\lceil \frac{2K}{N} \right\rceil \cdot \left( M + \frac{3N}{2} - 1\right).
\end{equation}

Since in the redundancy execution modes, several PEs calculate the same values, the term \textbf{effective size} of the systolic array is introduced. It defines the number of groups calculating unique values and the size of the output matrix. The effective size of the systolic array in the DRG execution mode is $N \times \frac{N}{2}$.

\begin{figure}[h]
    \centering
    \includegraphics[width=0.95\linewidth]{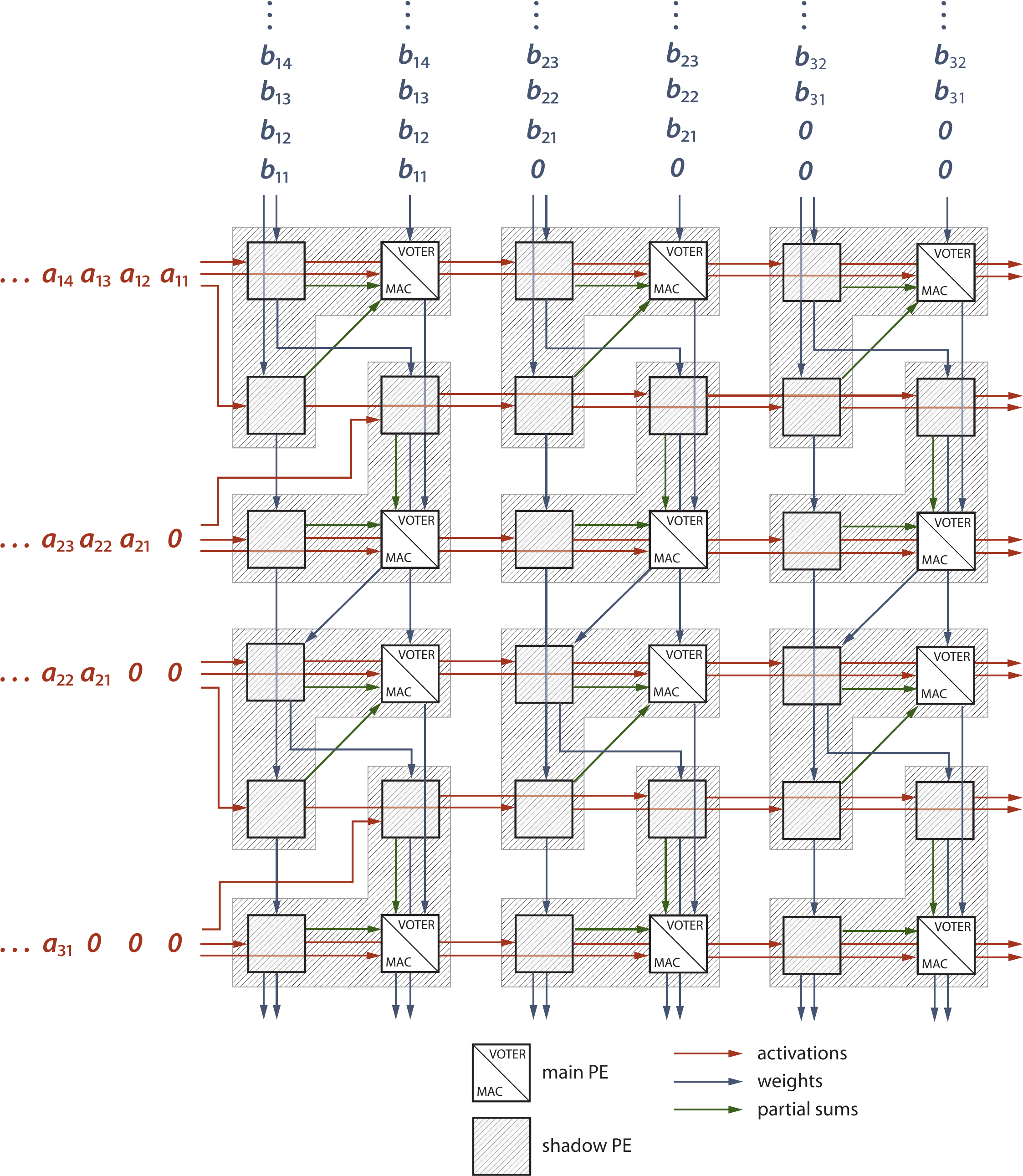}
    \caption{Arrangement of the systolic array in the triple-redundancy grouping execution mode, implementation option with groups of three PEs (TRG3)}
    \label{fig:tmr3_mode}
\end{figure}

\begin{figure}[h]
    \centering
    \includegraphics[width=0.95\linewidth]{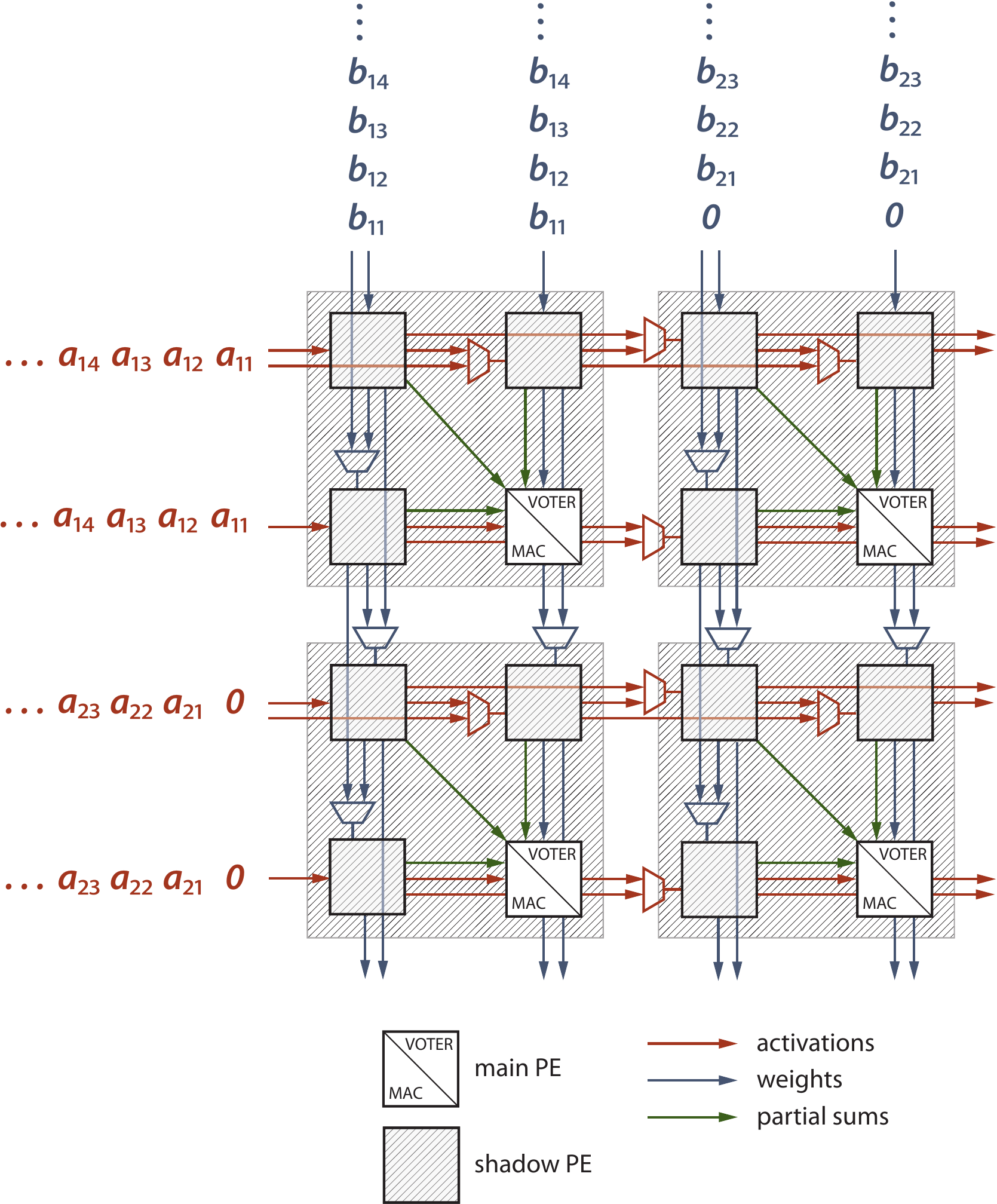}
    \caption{Arrangement of the systolic array in the triple-redundancy grouping execution mode, implementation option with groups of four PEs (TRG4)}
    \label{fig:tmr4_mode}
\end{figure}

\textbf{TRG execution mode.} For the TRG execution mode, two implementations are possible: with groups of three or four PEs. The implementation option with groups of three PEs (TRG3) is shown in Fig.~\ref{fig:tmr3_mode} and the implementation option with groups of four PEs (TRG4) in Fig.~\ref{fig:tmr4_mode}. The group size is the design-time parameter.

In the TRG3 implementation, the voter in the main PE works in parallel with the MAC unit. In the TRG4 implementation, the MAC unit in the main PE is not used in the TRG execution mode, the main PE only compares the partial sums calculated by three shadow PEs.

The effective size of the systolic array, i.e., the size of the output matrix, for TRG3 implementation is $\frac{2N}{3}\times\frac{N}{2}$. The effective sizes of the matrices that can be multiplied are $\frac{2N}{3}\times M$ and $M\times\frac{N}{2}$. The latency of the TRG3 implementation is
\begin{equation}
    L_{SA}^{trg3} = M + \frac{2N}{3} - 1 + \frac{N}{2} - 1 + 1 = M + \frac{7N}{6} - 1\,.
\end{equation}

The total latency then
\begin{equation}
    L_{total}^{trg3} = \left\lceil \frac{3P}{2N} \right\rceil \cdot \left\lceil \frac{2K}{N} \right\rceil \cdot \left(M + \frac{7N}{6} - 1\right).
\end{equation}

The effective size of the systolic array for the TRG4 implementation is $\frac{N}{2}\times\frac{N}{2}$. The effective sizes of the matrices that can be multiplied are $\frac{N}{2}\times M$ and $M\times\frac{N}{2}$. The latency of the TRG4 implementation is, therefore,
\begin{equation}
    L_{SA}^{trg4} = M + \frac{N}{2} - 1 + \frac{N}{2} - 1 + 1 = M + N - 1\,.
\end{equation}

The total latency then
\begin{equation}
    L_{total}^{trg4} = \left\lceil \frac{2P}{N} \right\rceil \cdot \left\lceil \frac{2K}{N} \right\rceil \cdot (M + N - 1).
\end{equation}

Since both execution modes have design-time parameters, four implementation options are possible: PM-DRG0-TRG3, PM-DRG0-TRG4, PM-DRGA-TRG3, and PM-DRGA-TRG4. Their efficiency is evaluated in Section~\ref{evaluation}. The overview of the execution modes and implementation options is given in Table~\ref{tab:modes_overview}.

\begin{table}[htbp]
    \caption{Execution modes (run-time) and implementation options (design-time)}
    \footnotesize
    \begin{center}
    \begin{tabular}{ccc}
        \toprule
        \textbf{Execution mode} & \textbf{Implementation options} & \textbf{Effective size} \\
        \midrule
        Performance (PM) & Baseline SA & $N\times N$ \\
        \midrule
        \multirow{2}{*}{DRG} & DRGA & \multirow{2}{*}{$\frac{N}{2}\times N$} \\
         & DRG0 & \\
        \midrule
        \multirow{4}{*}{TRG} & \multirow{2}{*}{TRG3} & \multirow{2}{*}{$\frac{2N}{3}\times \frac{N}{2}$} \\
         & & \\
         & \multirow{2}{*}{TRG4} & \multirow{2}{*}{$\frac{N}{2}\times \frac{N}{2}$} \\
         & & \\
        \bottomrule
    \end{tabular}
    \label{tab:modes_overview}
    \end{center}
\end{table}

The architecture of FORTALESA is based on the output-stationary dataflow; however, other dataflows can be used to build reconfigurable systolic arrays following FORTALESA methodology. Processing elements can be grouped in a similar way, with similar interleaved connections. During the pre-loading stage, PEs from one group should receive the same input or weight. The faults would be corrected by applying a selected correction technique to the output value.

\section{Reliability assessment method for systolic arrays}\label{reliability_assessment}
This section presents the proposed reliability assessment method for systolic arrays. The proposed reliability assessment method combines fault injection with fault propagation analysis. Instead of modeling a systolic array on the microarchitecture level and performing time-consuming cycle-accurate simulations, fault propagation analysis is used to calculate the resulting error in the layer output. This allows to noticeable speed up the reliability assessment of DNN inference on a systolic array. The method allows to analyze both transient and permanent faults in the OS systolic arrays.

\subsection{Fault propagation in systolic arrays}
\begin{figure}[h]
    \centering
    \includegraphics[width=0.9\linewidth]{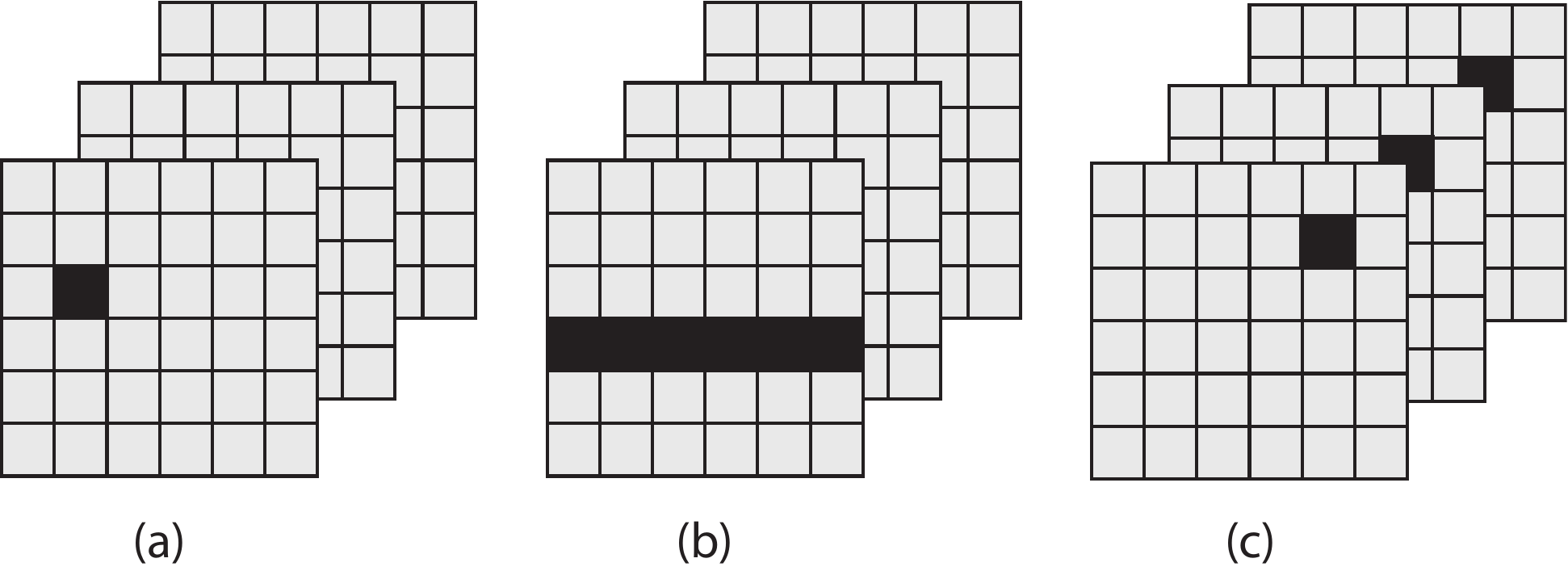}
    \caption{Error patterns in the output tensor: (a) single point, (b) line, (c) bullet}
    \label{fig:error_pattern}
\end{figure}

Considering the structure of a single processing element, faults can occur in registers holding intermediate values or in a multiplier. Considering the regular structure of a systolic array and data movement, faults in different components produce different error patterns in the output tensor. Fig. \ref{fig:error_pattern} presents the possible patterns for the convolution operation on the output-stationary systolic array. Fault patterns are derived analytically and experimentally following examples from previous works for similar cases \cite{Libano2023, Bolchini2023}. A bit flip in a weight register (WREG) results in the propagation of faulty weight through the whole column affecting several output values in one channel. This ends up in the \textit{line pattern}. A bit flip in an input register (IREG) results in the propagation of faulty activation through the whole row affecting one value in several output channels. This ends up in the \textit{bullet pattern}. A bit flip in the multiplier or an output register keeping partial sum (OREG) affects one value in one output channel, resulting in a \textit{point pattern}.

\subsection{Fault propagation analysis for transient faults}
Transient faults in registers and MAC units are considered for the analysis. Each fault is defined by seven parameters given in Table \ref{tab:fault_params_t}.

\begin{table}[htbp]
    \caption{Transient fault parameters}
    \footnotesize
    \begin{center}
    \begin{tabular}{rl}
        \toprule
        \textbf{Symbol} & \textbf{Description} \\
        \midrule
        $f_{type}$ & Type of fault (IREG, WREG, OREG, MULT) \\
        $ts$& Cycle of a systolic array execution \\
        $t_w$ & Weight tile \\
        $t_a$ & Activation tile \\
        $p_{row}$ & Row of the PE \\
        $p_{col}$ & Column of the PE \\
        $\beta_f$ & Bit index \\
        \bottomrule
    \end{tabular}
    \label{tab:fault_params_t}
    \end{center}
\end{table}

An output of the convolutional layer can be expressed as
\begin{equation}
    y_{(u,v)} = \sum_c^{C_{in}} \sum_i^{H_k} \sum_j^{W_k} w_{(c,i,j)} \cdot x_{(c,u+i,v+j)} + b\,,
\end{equation}
where $y_{(u,v)}$ is the output value with index $(u,v)$, $C_{in}$ is the number of input channels, $H_k$ and $W_k$ are the height and width of a kernel, $w_{(c,i,j)}$ is an individual weight, $x_{(c,u+i,v+j)}$ is an activation (input), and $b$ is a bias.

A bit flip in an input can be expressed as an addition of the error term $\varepsilon$ to this input. The error term for a signed integer is expressed as
\begin{equation}
    \varepsilon = 2^{\beta_f} \cdot \gamma\,,
\end{equation}
where $\beta_f$ is the index of a faulty bit and $\gamma$ is a sign coefficient, which is derived by the following equation
\begin{equation} \label{eq:sign_coeff}
    \gamma = \left\{ \begin{array}{rcl}
         -1 & \mbox{if} & x_f^{(\beta_f)} = 1 \wedge \beta_f \neq \beta_{sign} \\
         1  & \mbox{if} & x_f^{(\beta_f)} = 1 \wedge \beta_f = \beta_{sign} \\
         -1 & \mbox{if} & x_f^{(\beta_f)} = 0 \wedge \beta_f = \beta_{sign} \\
         1  & \mbox{if} & x_f^{(\beta_f)} = 0 \wedge \beta_f \neq \beta_{sign}
    \end{array} \right.
\end{equation}
where $x_f^{(\beta_f)}$ is the faulty bit of the affected input and $\beta_{sign}$ is the index of a sign bit.

A fault in a multiplier or register holding activation, weight, or partial sum results in an addition of an error $e$ to an output of the convolutional layer. This error $e$ depends on the error term $\varepsilon$ added to the faulty value and can be calculated for each type of fault from parameters given in Table \ref{tab:fault_params_t}.

\textbf{Fault in the input register of a systolic array.} If the fault happens in the input register, it corrupts an activation, then an error added to the output of the convolutional layer equals
\begin{equation}
    e_{ireg} = w_{(c_f,i_f,j_f)} \cdot \varepsilon\,, \label{eq:e_ireg}
\end{equation}
where $w_{(c_f,i_f,j_f)}$ is the weight multiplied with the faulty activation. This weight is found by its position $(c_f,i_f,j_f)$:
\begin{align}
    c_f &= \left\lceil \frac{ts - p_{col} - p_{row}}{H_k \cdot W_k}\right\rceil \\
    k_f &= (ts - p_{col} - p_{row}) \mod (H_k \cdot W_k) \\
    i_f &= \left\lfloor \frac{k_f}{W_k} \right\rfloor \\
    j_f &= k_f \mod W_k
\end{align}
where $ts$ is the targeted cycle of a systolic array execution, $p_{row}$ and $p_{col}$ are the position of the targeted PE.

Fault results in the bullet error pattern. Therefore, the following parameters define affected output values: output channels $c_{out,f}$ and position in the feature map $(u_f,v_f)$. Affected output channels are calculated as follows:
\begin{align}
    c_{out,f} &= \left[c_{out,f}^{start}, c_{out,f}^{end} \right] \\
    c_{out,f}^{start} &= (t_w - 1) \cdot N + p_{col} + 1 \label{eq:c_out_s}\\
    c_{out,f}^{end} &= \left\{ \begin{array}{rcl}
         t_w \cdot N & \mbox{if} & C_{out} \geqslant t_w \cdot N \\
         C_{out}  & \mbox{if} & C_{out} < t_w \cdot N
    \end{array} \right. \label{eq:c_out_e}
\end{align}
where $c_{out,f}$ is a range of affected output channels, starting with $c_{out,f}^{start}$ and ending with $c_{out,f}^{end}$, $t_w$ is the targeted weight tile, $p_{row}$ is the row of the targeted PE, and $C_{out}$ is the total number of output channels.

The position $(u_f,v_f)$ is found in the following way:
\begin{align}
    \Box_f &= (t_a - 1) \cdot N + p_{row} \label{eq:sw_f}\\
    u_f &= \left\lfloor \frac{\Box_f}{W_{out}} \right\rfloor \label{eq:u_f} \\
    v_f &= \Box_f \mod W_{out} \label{eq:v_f}
\end{align}
where $\Box_f$ is the affected sliding window and $W_{out}$ is the width of the output feature map.

An error $e_{ireg}$ calculated using equation \eqref{eq:e_ireg} is added to the affected output values.

\textbf{Fault in the weight register of a systolic array.} An error added to the output of the convolutional layer in this case equals
\begin{equation}
    e_{wreg} = x_{(c_f,i_f,j_f)} \cdot \varepsilon\,.
\end{equation}
The sign of the $\varepsilon$ is defined by the affected weight following equation \eqref{eq:sign_coeff}.

Fault in the weight register results in the line error pattern, therefore, the following parameters define affected output values: output channel $c_{out,f}$ and a range of positions starting with $(u_f^{start},v_f^{start})$ and ending with $(u_f^{end},v_f^{end})$. The affected output channel is calculated as follows:
\begin{equation}
    c_{out,f} = (t_w - 1) \cdot N + p_{col} \label{eq:c_out}\,.
\end{equation}

Positions are calculated from the affected sliding windows following equations \eqref{eq:u_f} and \eqref{eq:v_f}:
\begin{align}
    \Box_f^{start} &= (t_a - 1) \cdot N + p_{row} \\
    \Box_f^{end} &= \left\{ \begin{array}{rcl}
         t_a \cdot N & \mbox{if} & W_{out} \cdot H_{out} \geqslant t_a \cdot N \\
         W_{out} \cdot H_{out}  & \mbox{if} & W_{out} \cdot H_{out} < t_a \cdot N
         \end{array} \right.
\end{align}

\textbf{Faults in the output register and the multiplier.} Since those faults result in a single point error pattern, an error added to the output equals
\begin{equation}
    e_{oreg} = e_{mult} = 2^{\beta_f}\,.
\end{equation}

The output channel $c_{out,f}$ is found using equation \eqref{eq:c_out}, and the position is found using equations \eqref{eq:u_f} and \eqref{eq:v_f}.

\subsection{Fault propagation analysis for permanent faults}
Permanent faults are present during the whole execution and therefore defined by four parameters given in Table \ref{tab:fault_params_p}.

\begin{table}[htbp]
    \caption{Permanent fault parameters}
    \footnotesize
    \begin{center}
    \begin{tabular}{rl}
        \toprule
        \textbf{Symbol} & \textbf{Description} \\
        \midrule
        $f_{type}$ & Type of fault (IREG, WREG, OREG, MULT) \\
        $p_{row}$ & Row of the PE \\
        $p_{col}$ & Column of the PE \\
        $\beta_f$ & Bit index \\
        \bottomrule
    \end{tabular}
    \label{tab:fault_params_p}
    \end{center}
\end{table}

Permanent faults result in several error patterns, one for each step. Therefore, the affected output values are defined by sets of parameters. E.g., in case of permanent faults in the input registers, the affected output values are defined by a set of output channels
\begin{equation}
    \left\{ (c_{out,f}^{start},c_{out,f}^{end})_i \ | \ i \in \{1,...,T_w\} \right\}
\end{equation}
and a set of positions
\begin{equation}
    \left\{ (u_f,v_f)_i \ | \ i \in \{1,...,T_a\} \right\}.
\end{equation}

A set of output channels is found by substituting $t_w$ with an iterator in equations \eqref{eq:c_out_s} and \eqref{eq:c_out_e}:
\begin{align}
    c_{out,f(i)}^{start} &= (i - 1) \cdot N + p_{col} + 1 \ \forall \ i \in \{1,...,T_w\}\\
    c_{out,f(i)}^{end} &= \left\{ \begin{array}{rcl}
         i \cdot N & \mbox{if} & C_{out} \geqslant i \cdot N \\
         C_{out}  & \mbox{if} & C_{out} < i \cdot N
    \end{array} \right. \ \forall \ i \in \{1,...,T_w\}
\end{align}
which gives a pair $(c_{out,f}^{start},c_{out,f}^{end})$ for every weight tile.

A set of positions is found by substituting $t_a$ with an iterator in equation \eqref{eq:sw_f} for the affected sliding windows and then using equations \eqref{eq:u_f} and \eqref{eq:v_f} to calculate position from the sliding window:
\begin{align}
    \Box_{f(i)} &= (i - 1) \cdot N + p_{row} \ \forall \ i \in \{1,...,T_a\}\\
    u_{f(i)} &= \left\lfloor \frac{\Box_{f(i)}}{W_{out}} \right\rfloor \label{eq:u_fp}\\
    v_{f(i)} &= \Box_{f(i)} \mod W_{out} \label{eq:v_fp}
\end{align}
which gives a position for every input tile. This way, every bullet pattern is characterized by a Cartesian product of those two sets.

An error added to the output for each bullet pattern is the cumulative value
\begin{equation}
    e_{ireg}^p = \sum_c^{C_{in}} \sum_i^{H_k} \sum_j^{W_k} w_{(c,i,j)} \cdot \varepsilon_{(c,u_f+i,v_f+j)}
\end{equation}
where $\varepsilon_{(c,u_f+i,v_f+j)}$ is the error term added to input $x_{(c,u_f+i,v_f+j)}$. Since permanent faults are modeled as stuck-at-0 or stuck-at-1 faults, an error term is expressed as
\begin{equation}
    \varepsilon = \left\{ \begin{array}{lcl}
        2^{\beta_f} \cdot \gamma & \mbox{if} &  x^{(\beta_f)}_f \neq s \\
        0 & \mbox{if} & x^{(\beta_f)}_f = s
    \end{array}
    \right.
\end{equation}
where $s$ is the stuck-at state (0 or 1).

\subsection{Implementation}
\begin{figure}[h]
    \centering
    \includegraphics[width=0.95\linewidth]{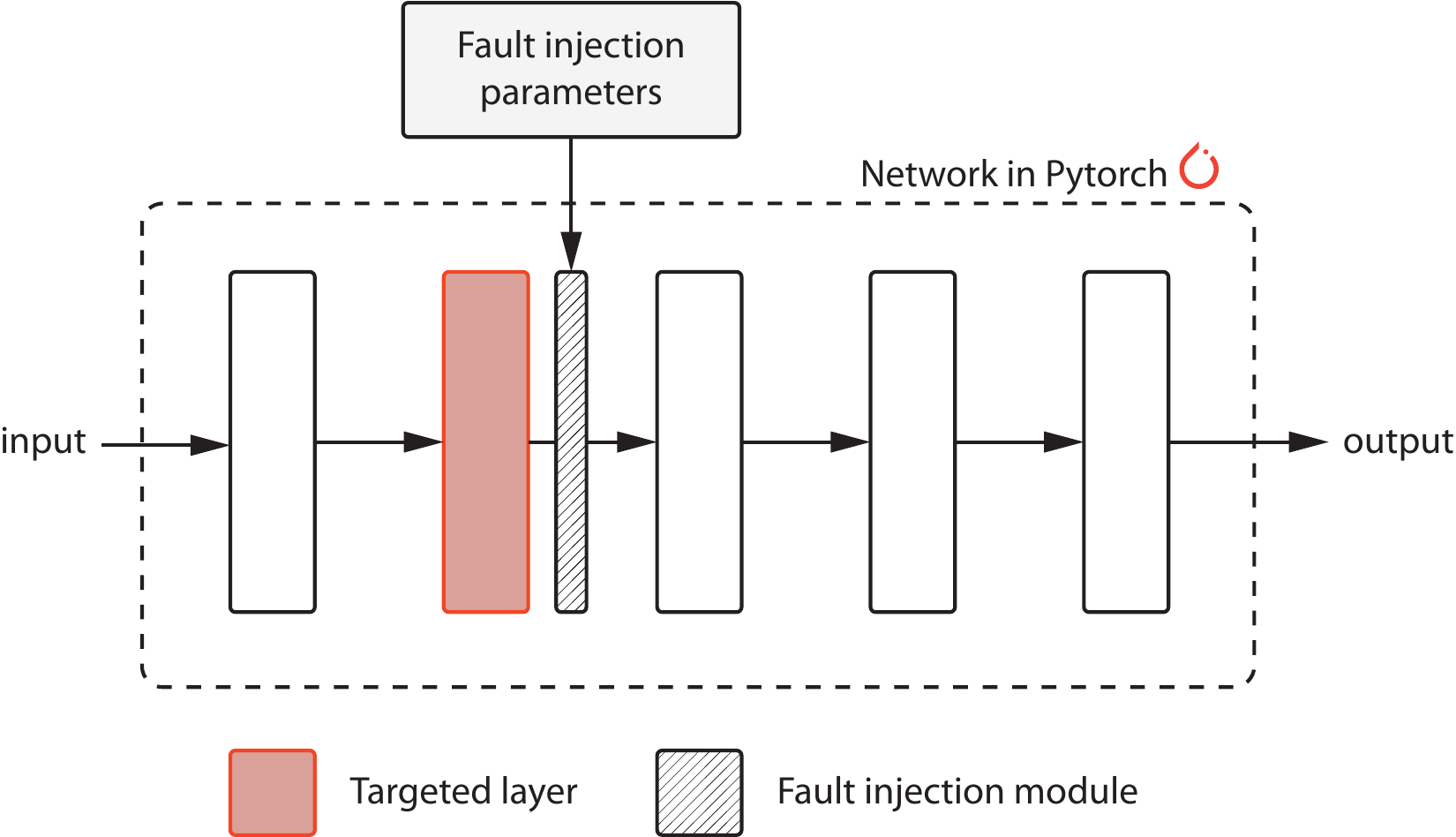}
    \caption{Workflow of the fault injection process}
    \label{fig:fi_workflow}
\end{figure}

The workflow of the proposed reliability assessment method combining fault injection with fault propagation analysis is presented in Fig. \ref{fig:fi_workflow}. Pytorch is used to run the inference of the network, leveraging GPU speed up. The fault injection module takes the output of the targeted layer and modifies only affected output values according to the aforementioned formulas. The simulation then continues with the erroneous values. The implementation of the proposed reliability assessment method will be made open-source.

\begin{figure*}[h]
    \centering
    \includegraphics[width=0.75\textwidth]{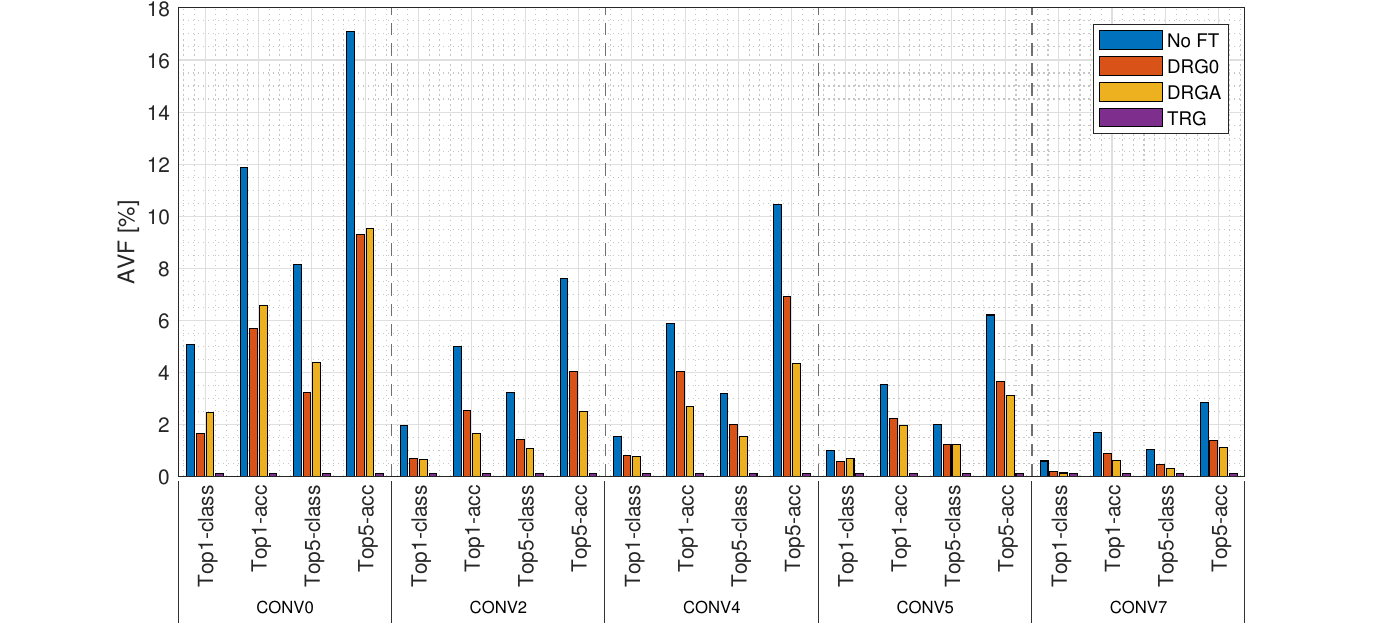}
    \caption{Layer-wise reliability assessment of AlexNet considering transient faults}
    \label{fig:alexnet_avf}
\end{figure*}

\begin{figure*}[h]
    \centering
    \includegraphics[width=0.95\textwidth]{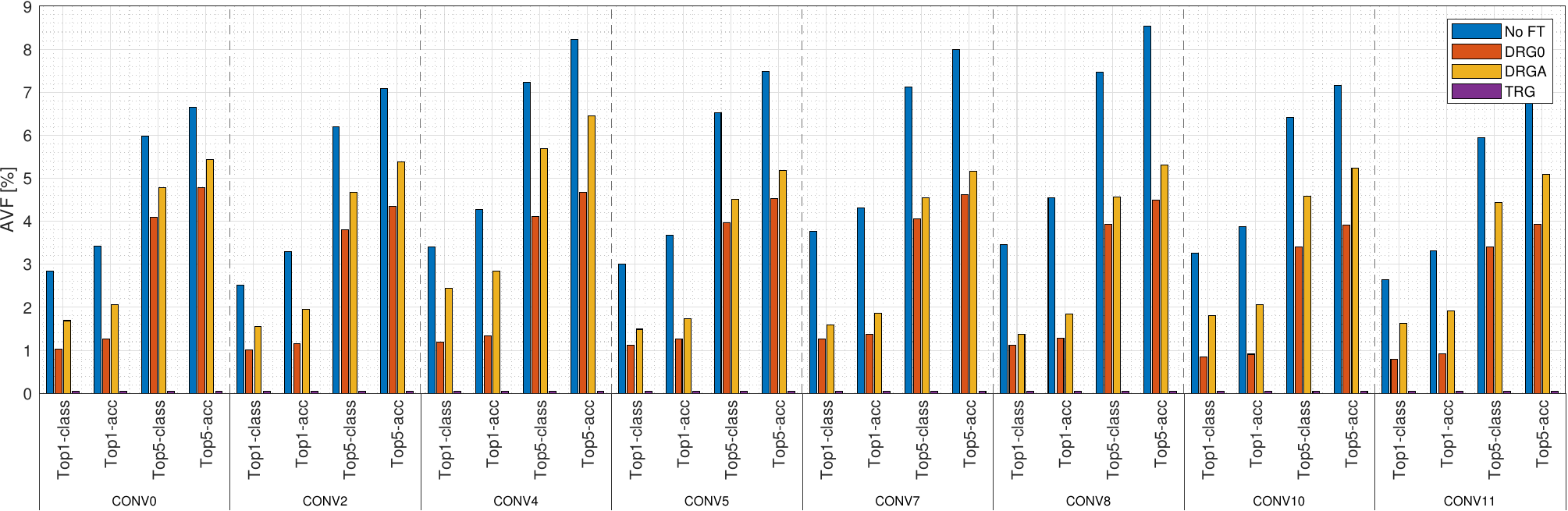}
    \caption{Layer-wise reliability assessment of VGG-11 considering transient faults}
    \label{fig:vgg_avf}
\end{figure*}

\begin{figure*}
    \centering
    \begin{subfigure}[b]{0.99\textwidth}
        \centering
        \includegraphics[width=\textwidth]{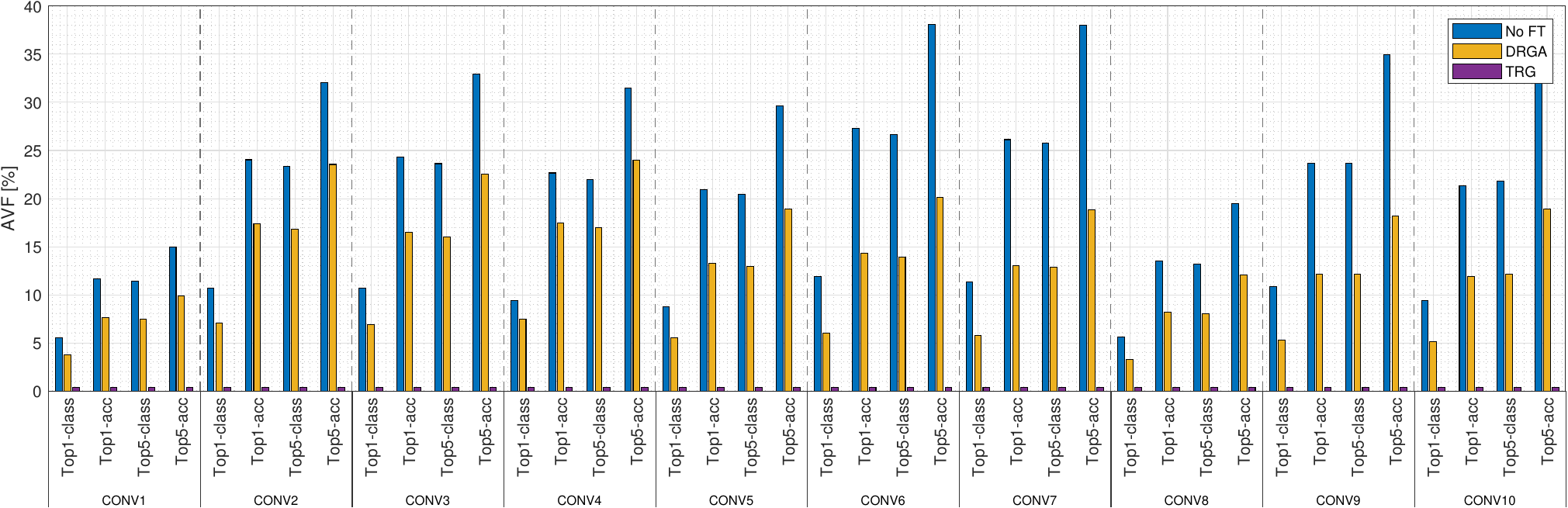}
        \vspace{1pt}
    \end{subfigure}
    \begin{subfigure}[b]{0.99\textwidth}
        \centering
        \includegraphics[width=\textwidth]{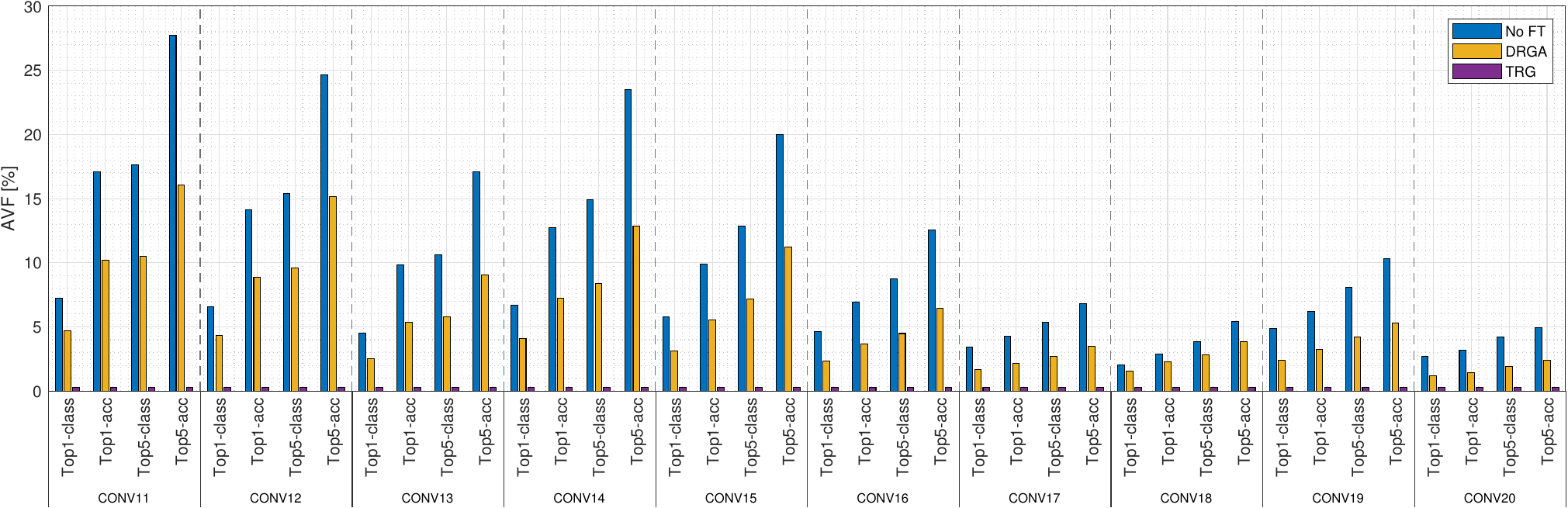}
    \end{subfigure}   
    \caption{Layer-wise reliability assessment of ResNet-18 considering transient faults}
    \label{fig:resnet_avf}
\end{figure*}

\section{Evaluation}\label{evaluation}
\subsection{Efficiency of reliability assessment method}
The proposed reliability assessment method for systolic arrays is compared with the representative fault injection framework that models systolic array microarchitecture, Saca-FI \cite{Tan2023}. While previous works rely on a cycle-accurate simulation of systolic array, which is a very time-consuming task, the proposed method is based on fault propagation analysis. Table \ref{tab:exec_time} presents the comparison of two fault injection methods. Runtime denotes the processing time of one input with fault injection. For runtime comparison, the selected model is VGG-16 evaluated on the ILSVRC-2012 dataset, systolic array size is 256$\times$256, fault is injected in the first convolutional layer. Both methods give AVF (Architecture Vulnerability Factor) as an output, the probability that a fault in the hardware structure causes an application output error \cite{Mukherjee2003}. While the functionality of the methods is similar, the proposed method is several orders of magnitude faster than the microarchitecture-level fault injection.

\begin{table}[htbp]
    \caption{Comparison of fault injection methods}
    \footnotesize
    \begin{center}
    \begin{tabular}{rp{2.5cm}ccc}
        \toprule 
        \textbf{Work} & \textbf{Fault model} & \textbf{Output} & \textbf{Dataflow} & \textbf{Runtime} \\
        \midrule
        Saca-FI & Transient and permanent faults in registers & AVF & WS, IS, OS & 47 s\\
        Ours & Transient and permanent faults in registers and multipliers & AVF & OS* & 0.0028 s\\
        \bottomrule
        \multicolumn{5}{l}{} \\
        \multicolumn{5}{l}{* The proposed method can be easily extended to other dataflows as well.} \\
    \end{tabular}
    \label{tab:exec_time}
    \end{center}
    \vspace{-4mm}
\end{table}

\subsection{FORTALESA hardware implementation parameters}
The proposed architecture of the systolic array is described in SystemVerilog and synthesized using Cadence Genus 2021 and 45nm Nangate PDK. A comparison of the baseline systolic array without reconfiguration ability and four implementation options of FORTALESA is presented in Tables \ref{tab:hw_params_48} and \ref{tab:hw_params_132}. Two sizes of the systolic array were selected: 48$\times$48 and 132$\times$132 to demonstrate the scalability of the design. Input and weight registers are 8-bit long, and the output register is 32-bit long. Due to the regular structure of the proposed architecture, the designs scale almost linearly proportionate to the increase in the number of processing elements. All implementations introduce around 5\% overhead due to additional interconnects and MUXes, around 6-7\% due to correction mechanisms for DRG0 versions, and around 17-18\% for DRGA versions.

It can be noticed that frequency might drop considerably for the implementations with the averaging correction scheme for the DRG execution mode (DRGA). Critical path of these implementation variants includes an additional adder in the main PEs, thus increasing the datapath and lowering the frequency.

\begin{table}[htbp]
    \caption{Hardware implementation parameters for 48$\times$48 systolic array}
    \footnotesize
    \begin{center}
    \begin{tabular}{rccc}
        \toprule 
        \textbf{Implementations} & \textbf{Area, mm$^2$} & \textbf{Power, W} & \textbf{Max frequency, MHz} \\
        \midrule
        Baseline SA & 1.726 & 0.158 & 402\\
        PM-DRG0-TRG3 & 1.937 & 0.177 & 357\\
        PM-DRG0-TRG4 & 1.929 & 0.176 & 372\\
        PM-DRGA-TRG3 & 2.129 & 0.193 & 303\\
        PM-DRGA-TRG4 & 2.091 & 0.190 & 302\\
        \bottomrule
    \end{tabular}
    \label{tab:hw_params_48}
    \end{center}
    \vspace{-4mm}
\end{table}

\begin{table}[htbp]
    \caption{Hardware implementation parameters for 132$\times$132 systolic array}
    \footnotesize
    \begin{center}
    \begin{tabular}{rccc}
        \toprule
        \textbf{Implementations} & \textbf{Area, mm$^2$} & \textbf{Power, W} & \textbf{Max frequency, MHz} \\
        \midrule
        Baseline SA & 12.533 & 3.625 & 285\\
        PM-DRG0-TRG3 & 14.243 & 4.053 & 252\\
        PM-DRG0-TRG4 & 14.284 & 4.029 & 249\\
        PM-DRGA-TRG3 & 15.826 & 4.552 & 230\\
        PM-DRGA-TRG4 & 15.605 & 4.489 & 262\\
        \bottomrule
    \end{tabular}
    \label{tab:hw_params_132}
    \end{center}
    \vspace{-4mm}
\end{table}

\subsection{Reliability analysis}
\textbf{Experimental setup.} Three common CNN models were used for reliability evaluation, including AlexNet, trained on the CIFAR-10 dataset, and VGG-11 and ResNet-18, trained on the ILSVRC-2012 dataset. Models and datasets were selected to represent different network scales. For the fault injection experiments, networks were quantized to an 8-bit integer format. Experiments were performed on NVIDIA GeForce RTX 3090 24G GPU.

Architectural vulnerability factor (AVF) was used as a reliability measure. AVF is the probability that a fault in the hardware structure causes an application output error \cite{Mukherjee2003}. Since the output of the selected models is the probability score of each class, the following output errors of the DNNs were considered \cite{Tan2023}:
\begin{itemize}
    \item Top1-class: the top-ranked class is different from the golden run.
    \item Top1-acc: the probability score of the top-ranked class is different. The top-ranked class may be different. Top1-acc includes the Top1-class errors.
    \item Top5-class: at least one class in the top-5 is different, including different order of the top-5 classes.
    \item Top5-acc: the probability score of at least one class in the top-5 is different. The top-5 classes may be different. Top5-acc includes Top1-class, Top1-acc and Top5-class errors.
\end{itemize}

For transient fault analysis, layer-wise statistical fault injection was done following the equation introduced in \cite{Leveugle2009} to achieve 95\% confidence and 5\% error margin. For each fault injection, fault parameters (Table \ref{tab:fault_params_t}) were set randomly and a test dataset of 10.000 inputs was fed to the network. Faults were injected in each convolution layer using the proposed methodology (Section \ref{reliability_assessment}). For the TRG execution mode, it is assumed that all faults are corrected, and there are no output errors. Fully connected layers were not considered for fault injection as they occupy only one row of the systolic array and their vulnerability factor, therefore, is very low \cite{Tan2023}.

\textbf{Experimental results.} Figures \ref{fig:alexnet_avf}, \ref{fig:vgg_avf}, and \ref{fig:resnet_avf} show AVF values of each layer for selected networks considering different execution modes and configurations of the $48\times48$ systolic array and transient faults. For ResNet-18, only DRGA is considered since DRG0 correction scheme was not very efficient for this model.

\begin{figure}[h]
    \centering
    \includegraphics[width=0.95\linewidth]{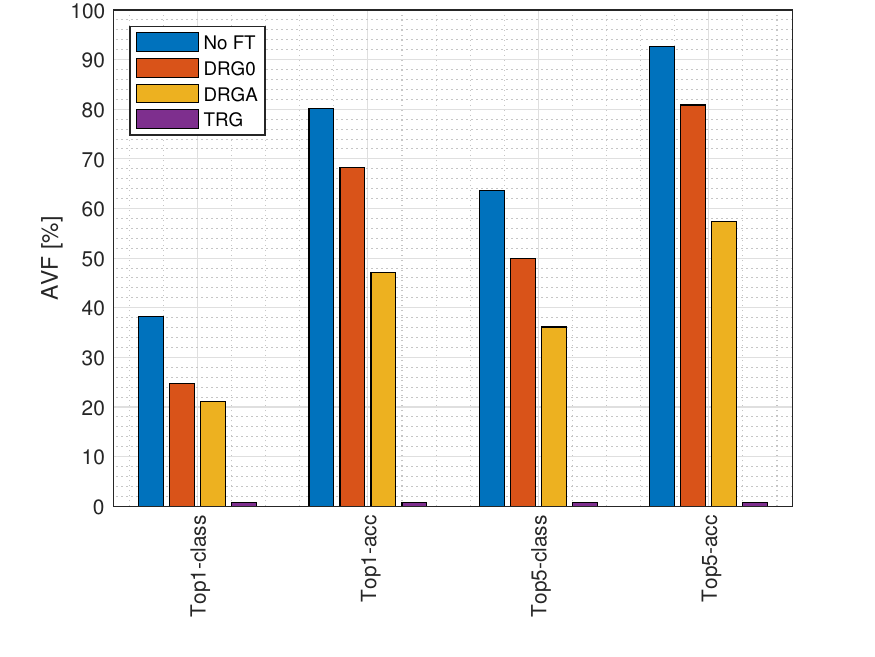}
    \caption{Reliability assessment of AlexNet considering permanent faults (stuck-at-1)}
    \label{fig:alexnet_sa}
\end{figure}

Results show the varying vulnerability of different layers and the masking effect of DRG execution mode implementations. It can be seen that the DRG execution mode decreases the vulnerability of certain layers almost twice. Those results can be used to decide how to assign different layers of DNNs to different execution modes of the systolic array.

Figure \ref{fig:alexnet_sa} shows AVF values for the reliability assessment of permanent faults for AlexNet. The analysis is done for stuck-at-1 faults since they are proven to be more critical \cite{Tan2023}. Unlike transient faults, analysis is done for the whole network (all convolutional layers) as permanent faults persist throughout the network execution.

\subsection{Performance vs. reliability trade-off}
\begin{figure}
     \centering
     \begin{subfigure}[b]{0.24\textwidth}
         \centering
         \includegraphics[width=\textwidth]{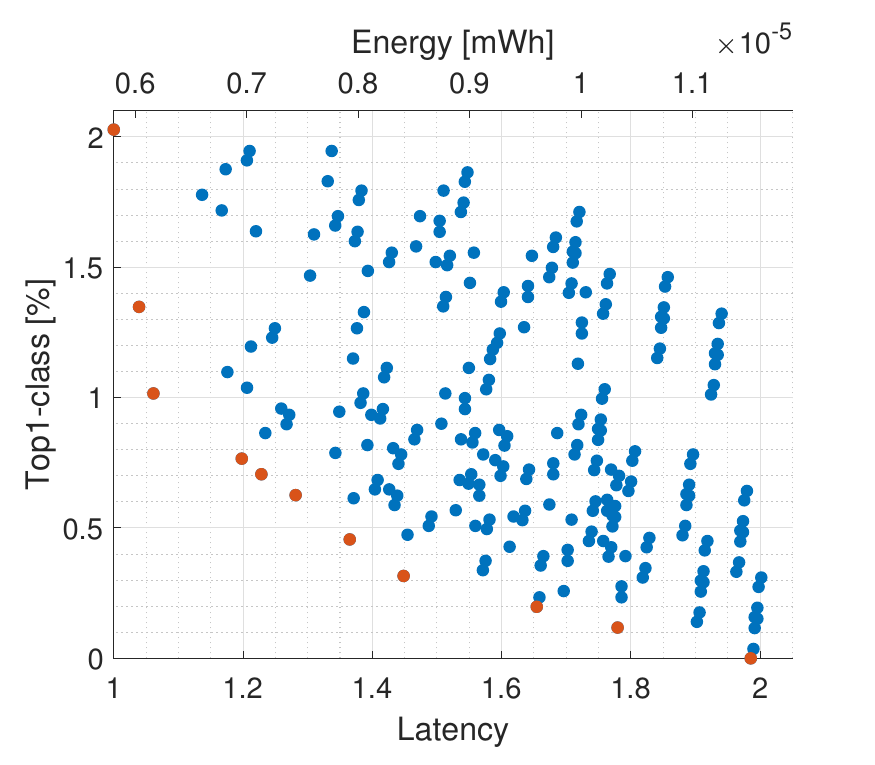}
         \caption{PM-DRG0-TRG3}
     \end{subfigure}
     \begin{subfigure}[b]{0.24\textwidth}
         \centering
         \includegraphics[width=\textwidth]{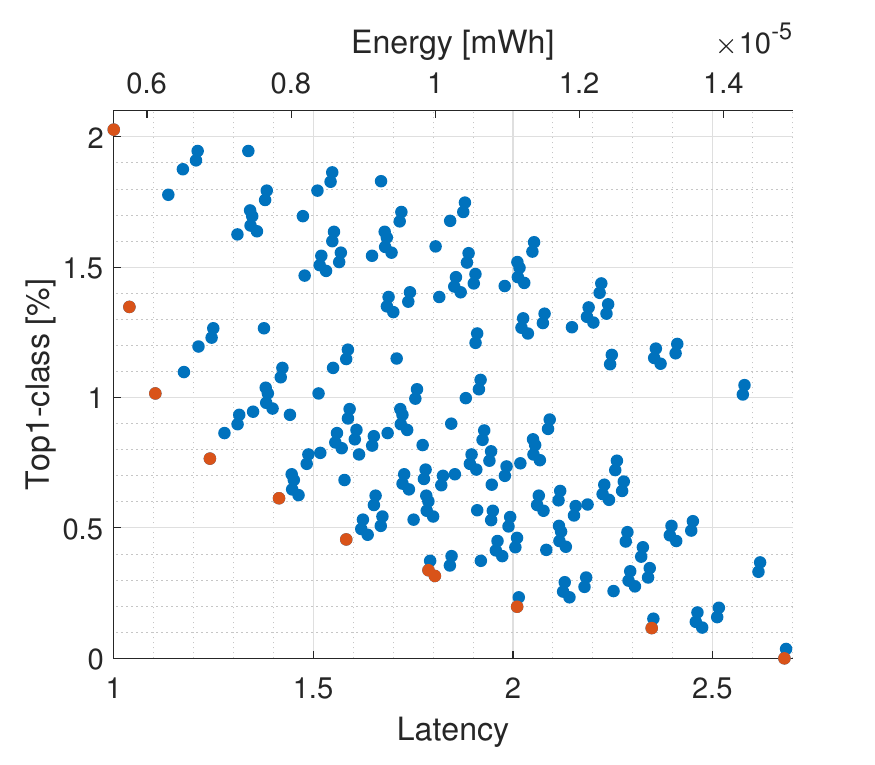}
         \caption{PM-DRG0-TRG4}
     \end{subfigure}
     \begin{subfigure}[b]{0.24\textwidth}
         \centering
         \includegraphics[width=\textwidth]{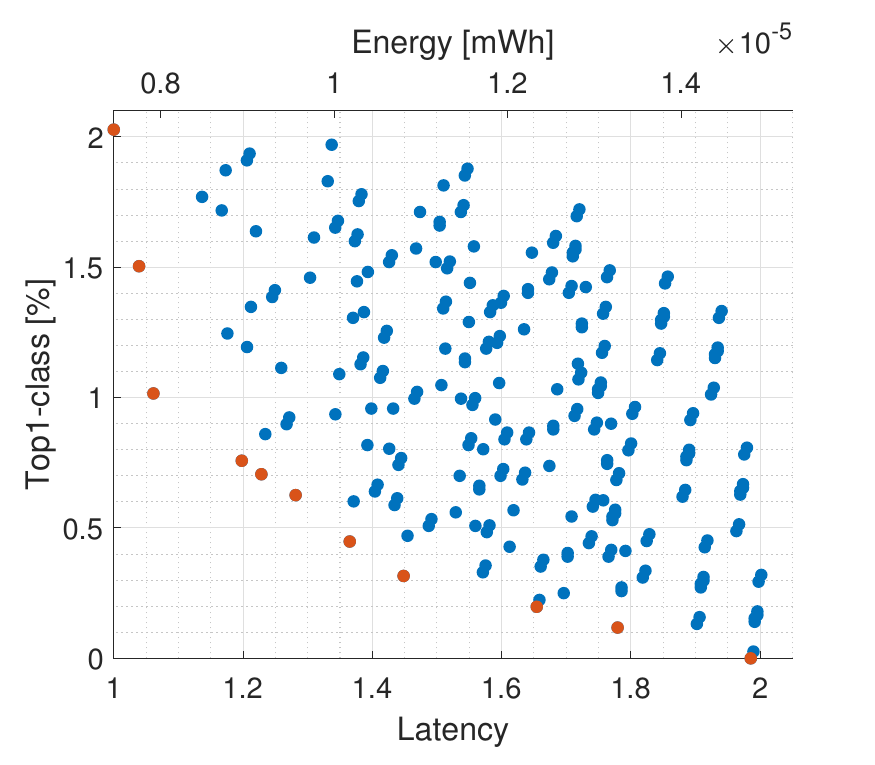}
         \caption{PM-DRGA-TRG3}
     \end{subfigure}
     \begin{subfigure}[b]{0.24\textwidth}
         \centering
         \includegraphics[width=\textwidth]{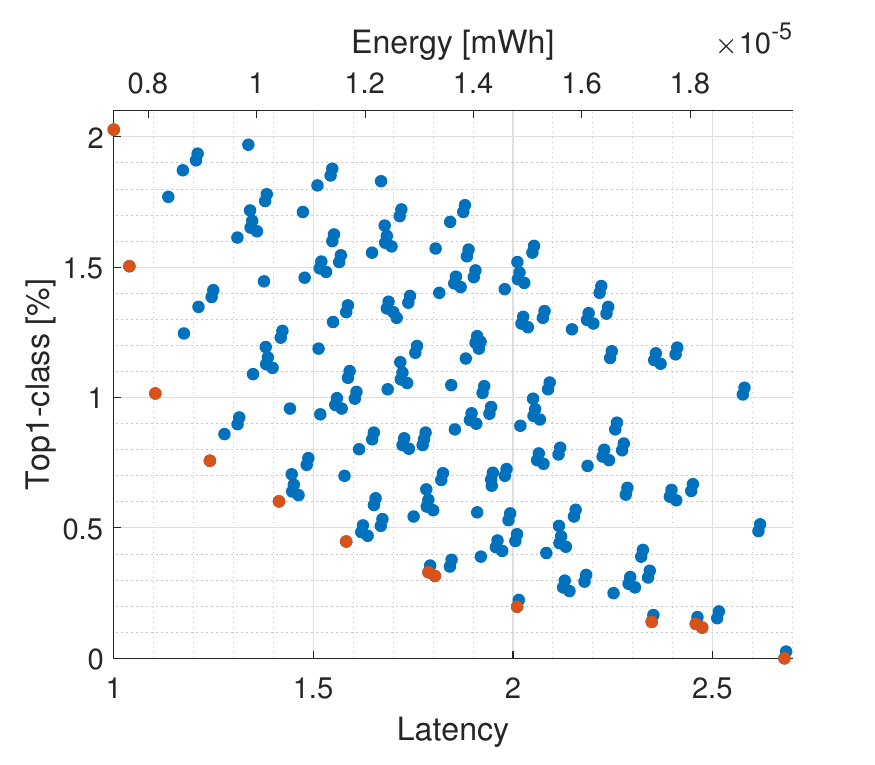}
         \caption{PM-DRGA-TRG4}
     \end{subfigure}
        \caption{Reliability vs. latency and energy for AlexNet}
        \label{fig:alexnet_configs}
\end{figure}

\begin{figure}
     \centering
     \begin{subfigure}[b]{0.24\textwidth}
         \centering
         \includegraphics[width=\textwidth]{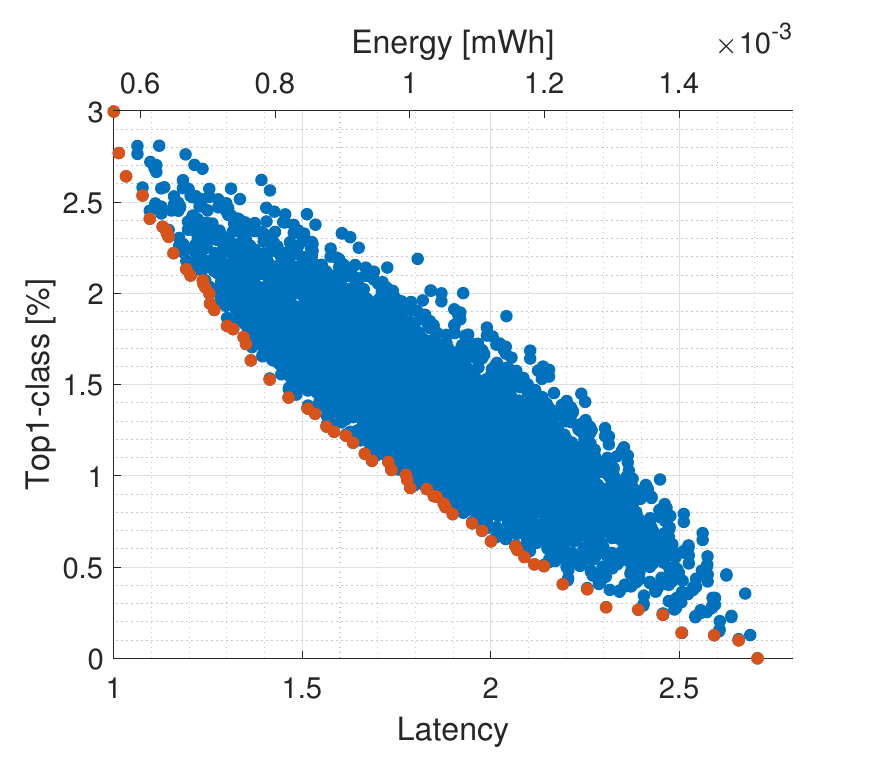}
         \caption{PM-DRG0-TRG3}
     \end{subfigure}
     \begin{subfigure}[b]{0.24\textwidth}
         \centering
         \includegraphics[width=\textwidth]{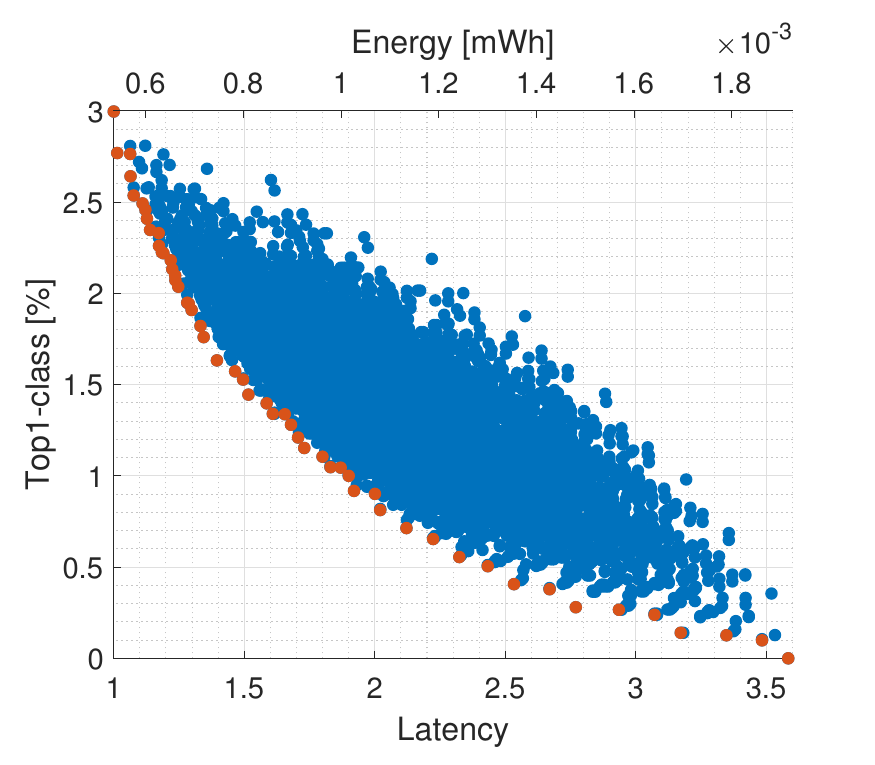}
         \caption{PM-DRG0-TRG4}
     \end{subfigure}
     \begin{subfigure}[b]{0.24\textwidth}
         \centering
         \includegraphics[width=\textwidth]{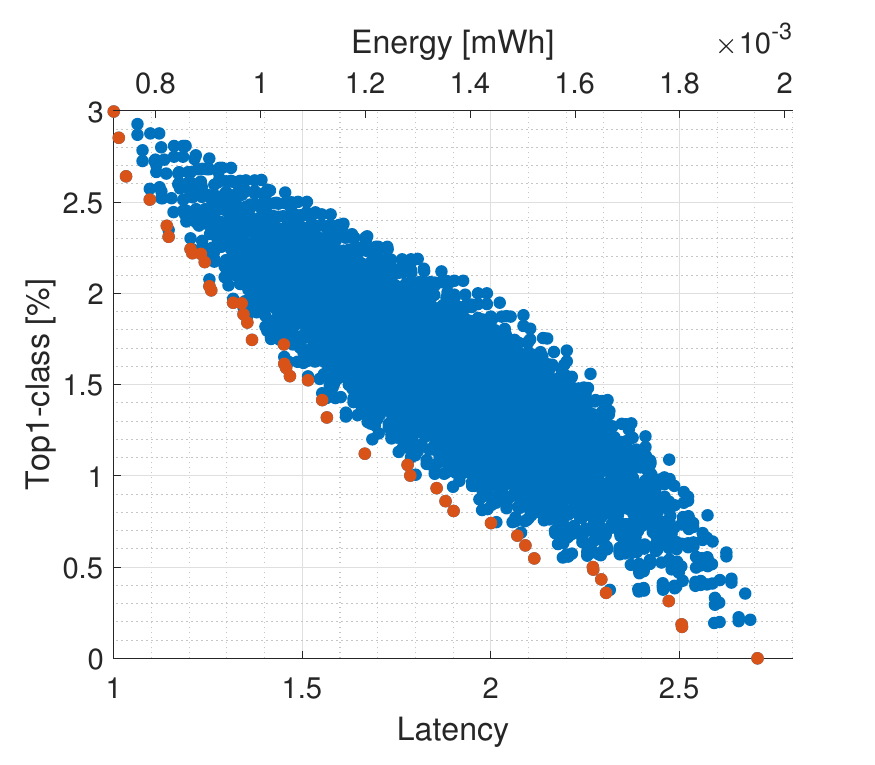}
         \caption{PM-DRGA-TRG3}
     \end{subfigure}
     \begin{subfigure}[b]{0.24\textwidth}
         \centering
         \includegraphics[width=\textwidth]{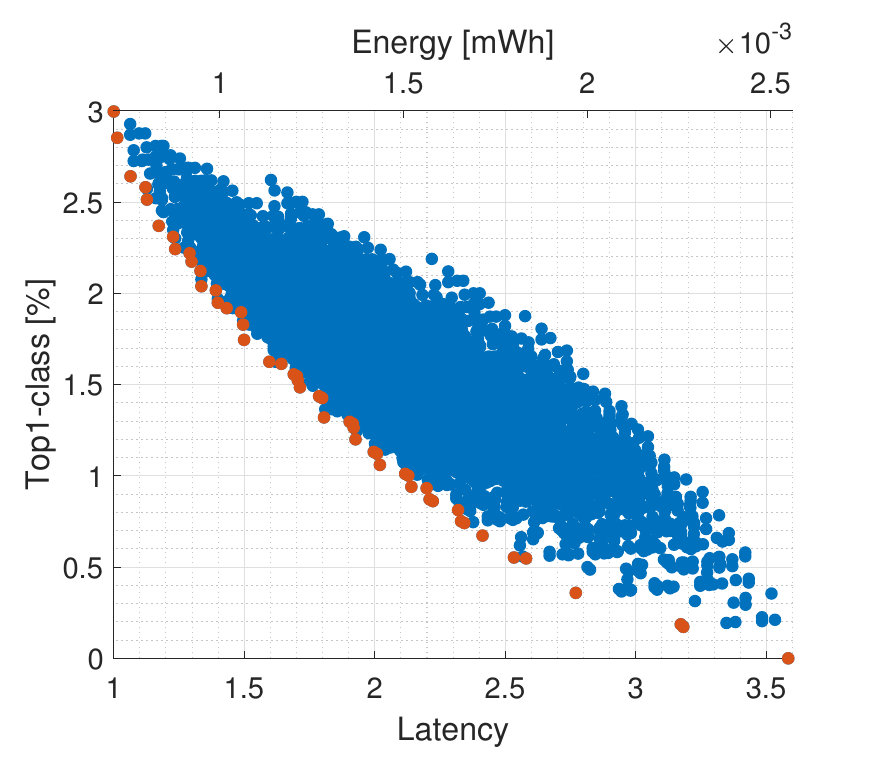}
         \caption{PM-DRGA-TRG4}
     \end{subfigure}
        \caption{Reliability vs. latency and energy for VGG-11}
        \label{fig:vgg_configs}
\end{figure}

\begin{figure}
    \centering
    \begin{subfigure}[b]{0.24\textwidth}
        \includegraphics[width=\textwidth]{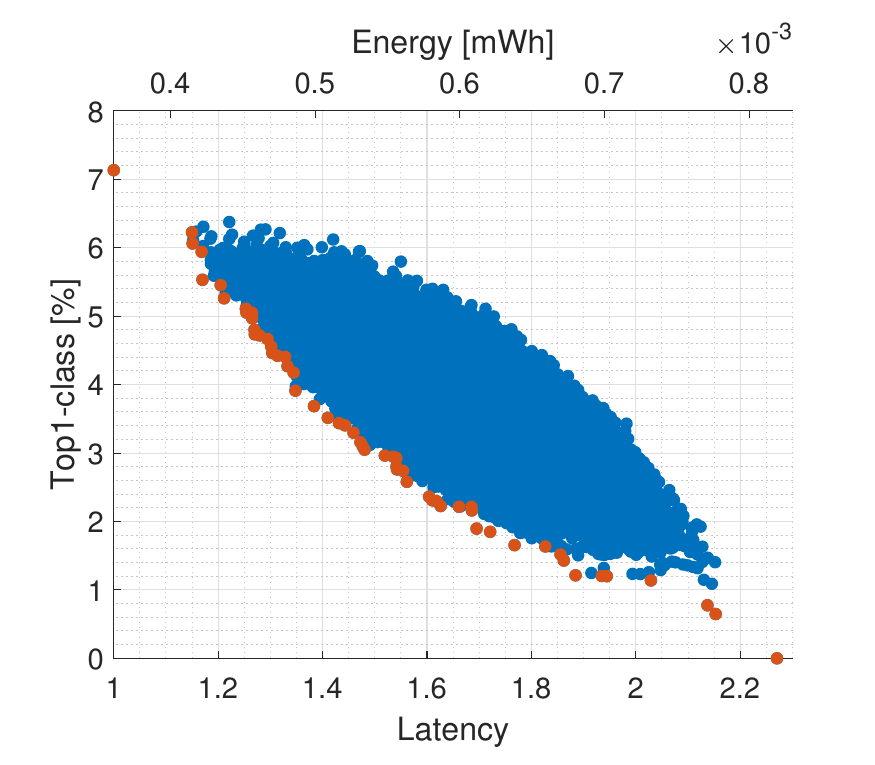}
        \caption{PM-DRGA-TRG3}
    \end{subfigure}
    \begin{subfigure}[b]{0.24\textwidth}
        \includegraphics[width=\textwidth]{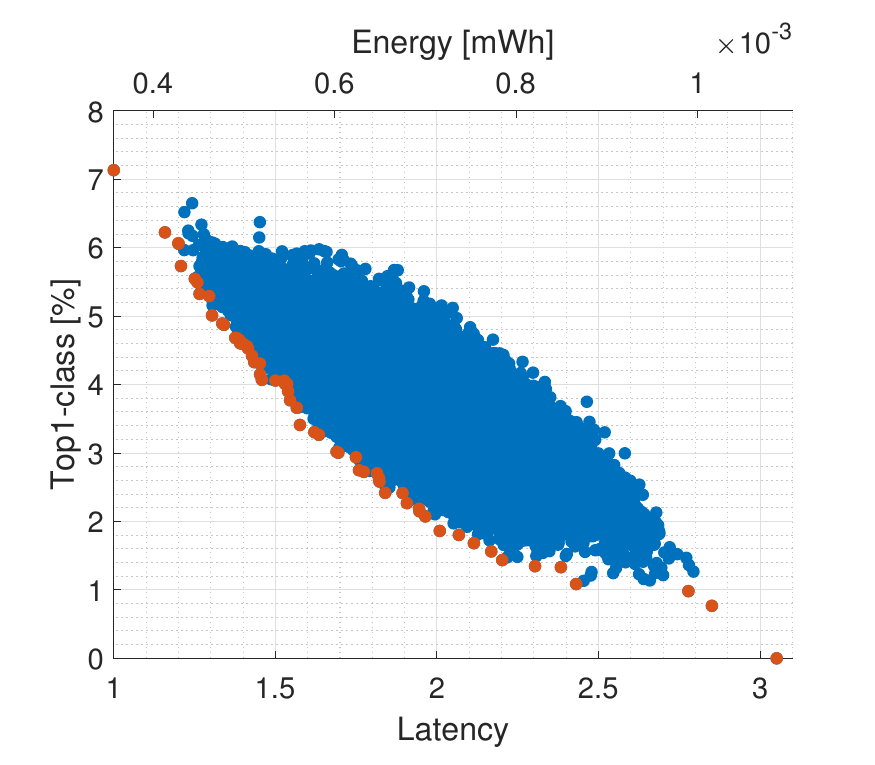}
        \caption{PM-DRGA-TRG4}
    \end{subfigure}
    \caption{Reliability vs. latency and energy for ResNet-18}
    \label{fig:resnet_configs}
\end{figure}

A flexible performance vs. reliability trade-off can be achieved by executing different layers of the DNN using different modes of the reconfigurable systolic array. Figures \ref{fig:alexnet_configs}, \ref{fig:vgg_configs} and \ref{fig:resnet_configs} plot reliability versus execution latency of the networks and energy consumption for all the possible combinations of modes. Latency is normalized to the execution of the network using performance mode for every layer (no fault tolerance). Energy is given in milliwatt-hour (mWh). For reliability, Top1-class errors are considered. The Pareto front is marked with red color. Four implementations of FORTALESA are considered for each network since different implementations of DRG execution mode give slightly different reliability results, and different implementations of TRG execution mode affect the latency.

The left-most point of the Pareto front denotes the configuration where all layers are executed using performance mode. The execution latency is small, but there is no fault tolerance. The right-most point denotes the configuration where all layers are executed using TRG mode. The execution latency is high, but all faults are corrected. The points in between denote configurations with a combination of execution modes, including DRG mode. The figures show that DRG execution mode allows for a flexible reliability--performance trade-off.

\begin{figure}[h]
    \centering
    \includegraphics[width=0.9\linewidth]{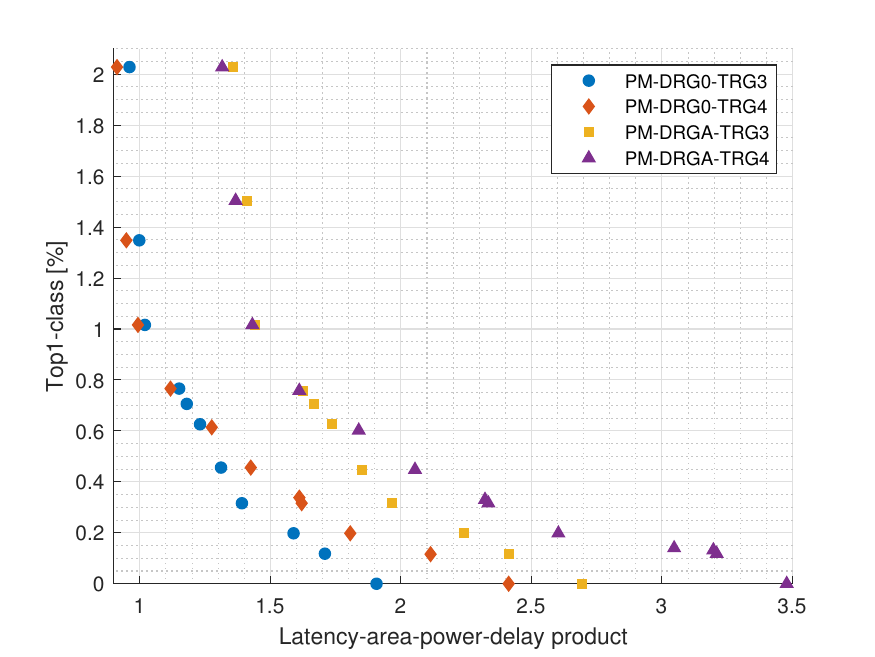}
    \caption{Comparison of FORTALESA implementation options considering reliability for AlexNet}
    \label{fig:alexnet_cfg_comp}
\end{figure}

\begin{figure}[h]
    \centering
    \includegraphics[width=0.9\linewidth]{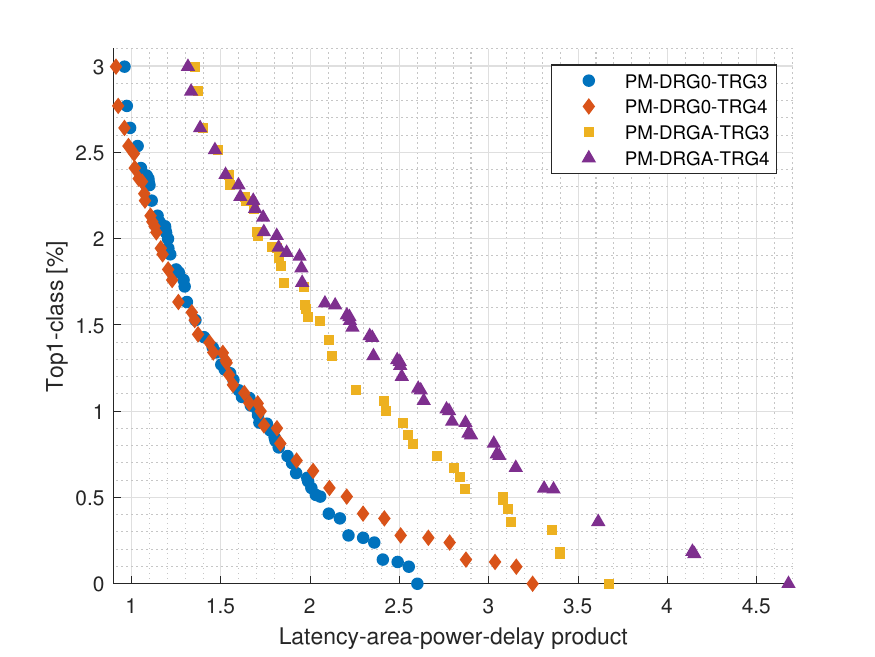}
    \caption{Comparison of FORTALESA implementation options considering reliability for VGG-11}
    \label{fig:vgg_cfg_comp}
\end{figure}

To compare different implementations of FORTALESA, the AVF of the Pareto front points was plotted against a product of latency, power, area, and delay (Fig. \ref{fig:alexnet_cfg_comp} and \ref{fig:vgg_cfg_comp}). It can be seen that while systolic arrays with the TRG4 implementation have smaller area and power, and higher frequency parameters, taking execution latency into consideration version with the TRG3 implementation shows better overall results.

These figures present a design space exploration with the highlighted Pareto front, allowing designers to select a trade-off that fits the needs and requirements of their system.

\subsection{Comparison with static redundancy}
\begin{figure}[h]
    \centering
    \includegraphics[width=0.9\linewidth]{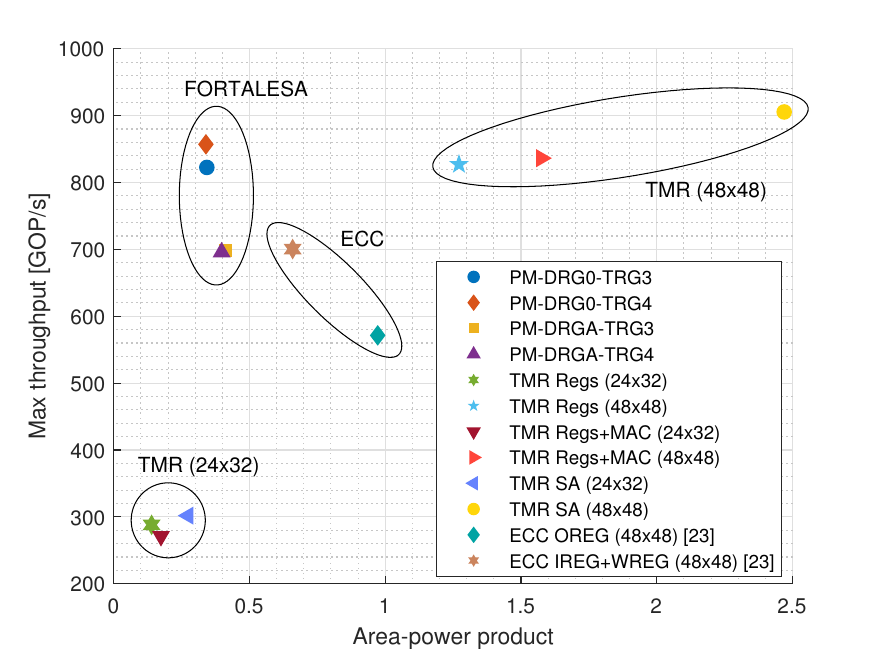}
    \caption{Comparison of FORTALESA implementation options with static TMR and selective ECC \cite{Tan2023}}
    \label{fig:tmr_comp}
\end{figure}

The comparison of the proposed run-time reconfigurable systolic array architecture with the static TMR approaches is presented in Fig. \ref{fig:tmr_comp}. A power-area product is plotted against the maximum possible throughput of the architecture (since the throughput of FORTALESA depends on the execution mode). Throughput is calculated as the number of MAC operations performed by the systolic array in one clock cycle multiplied by the frequency. Different cases are considered for static TMR: triplication of registers only, triplication of registers and MAC units, and triplication of the whole array. As well as different sizes: $48\times48$ and $24\times32$ as this is the efficient size of the $48\times48$ systolic array operating in a TRG execution mode (TRG3 implementation option). As can be seen from the figure, the $24\times32$ systolic array with static TMR has a lower power-area product than the proposed architecture but has a fixed low throughput. On the other hand, the $48\times48$ systolic array with static TMR has high throughput, but the power-area product is significantly higher than the proposed architecture. FORTALESA, on average, requires $6\times$ fewer resources than static TMR.

The proposed architecture is also compared with the selective ECC proposed in \cite{Tan2023}. While it protects only selected registers, it requires, on average, $2.5\times$ more resources than FORTALESA, which protects all registers and MAC units.

\subsection{Comparison with the state-of-the-art}
The comparison of the proposed architecture with other methods to enhance fault tolerance of systolic arrays is presented in Table \ref{tab:comp}. The proposed architecture provides a comprehensive solution that protects both registers and MAC units in the PEs, and since it utilizes a form of spatial redundancy, it covers both transient and permanent faults. Faults are corrected in both fault-tolerant modes without interrupting inference execution for recovery. Additionally, the proposed architecture provides support for reconfigurability, addressing the dynamic requirements of AI applications.

\begin{table}[htbp]
    \caption{Comparison with the state-of-the-art}
    \footnotesize
    \begin{center}
    \begin{tabular}{rccccc}
        \toprule 
        \multirow{2}{*}{\textbf{Work}} & \multicolumn{2}{c}{\textbf{Covered faults}} & \multicolumn{2}{c}{\textbf{Protected parts}} & \textbf{Fault} \\
         & \textbf{Permanent} & \textbf{Transient} & \textbf{Registers} & \textbf{MAC} & \textbf{correction}\\
        \midrule
        \cite{Zhao2022} & \ding{51} & \ding{55} & \ding{55} & \ding{51} & \ding{51} \\
        \cite{Zhang2019} & \ding{51} & \ding{55} & \ding{55} & \ding{51} & \ding{51}\textsuperscript{1} \\
        \cite{Libano2023} & \ding{51} & \ding{55} & \ding{55} & \ding{51}\textsuperscript{2} & \ding{55} \\
        \cite{Tan2023} & \ding{55} & \ding{51} & \ding{51} & \ding{55} & \ding{51}\textsuperscript{3} \\
        \cite{Vacca2023a} & \ding{51} & \ding{55} & \ding{51} & \ding{51} & \ding{55} \\
        \cite{Safarpour2022} & \ding{55} & \ding{51}\textsuperscript{4} & \ding{55} & \ding{51} & \ding{55} \\
        Our & \ding{51} & \ding{51} & \ding{51} & \ding{51} & \ding{51} \\
        \bottomrule
        \multicolumn{6}{l}{} \\
        \multicolumn{6}{l}{\textsuperscript{1} Requires retraining of the DNN for each faulty chip.} \\
        \multicolumn{6}{l}{\textsuperscript{2} Faults in configurational memory of FPGA.} \\
        \multicolumn{6}{l}{\textsuperscript{3} Only single-bit fault correction (ECC).} \\
        \multicolumn{6}{l}{\textsuperscript{4} Timing faults from low voltage.} \\
    \end{tabular}
    \label{tab:comp}
    \end{center}
\end{table}

\section{Conclusion}\label{conclusion}
In this work, we proposed a run-time reconfigurable fault-tolerant systolic array architecture FORTALESA with three execution modes and four implementation options. All four implementation options were evaluated in terms of resource utilization, throughput, and fault tolerance improvement. The proposed architecture is used for reliability enhancement of DNN inference on a systolic array through heterogeneous mapping of different network layers to different execution modes. We also introduced a reliability assessment method based on fault propagation analysis. It is used to efficiently evaluate the vulnerability of DNN models and select the appropriate execution mode--layer mapping for DNN inference. The proposed architecture efficiently protects both registers and MAC units of systolic array PEs from transient and permanent faults without interrupting inference execution. The reconfigurability feature of FORTALESA enables a speedup of up to $3\times$, depending on layer vulnerability, and requires $6\times$ fewer resources compared to static redundancy and $2.5\times$ fewer resources than the previously proposed solution for transient faults. As future work, an ASIC (Application-Specific Integrated Circuit) based on FORTALESA will be developed for further testing, e.g., at radiation facilities. The design of the chip will be made open-source.

\section*{Acknowledgment}
This work was supported in part by the Estonian Research
Council grant PUT PRG1467 ”CRASHLESS”, CoE TK202 "Universum" and by EU
Grant Project 101160182 “TAICHIP“.

\bibliographystyle{IEEEtran}
\bibliography{references}

\begin{thebibliography}{10}
\providecommand{\url}[1]{#1}
\csname url@samestyle\endcsname
\providecommand{\newblock}{\relax}
\providecommand{\bibinfo}[2]{#2}
\providecommand{\BIBentrySTDinterwordspacing}{\spaceskip=0pt\relax}
\providecommand{\BIBentryALTinterwordstretchfactor}{4}
\providecommand{\BIBentryALTinterwordspacing}{\spaceskip=\fontdimen2\font plus
\BIBentryALTinterwordstretchfactor\fontdimen3\font minus \fontdimen4\font\relax}
\providecommand{\BIBforeignlanguage}[2]{{%
\expandafter\ifx\csname l@#1\endcsname\relax
\typeout{** WARNING: IEEEtran.bst: No hyphenation pattern has been}%
\typeout{** loaded for the language `#1'. Using the pattern for}%
\typeout{** the default language instead.}%
\else
\language=\csname l@#1\endcsname
\fi
#2}}
\providecommand{\BIBdecl}{\relax}
\BIBdecl

\bibitem{TPU}
N.~P. Jouppi, C.~Young, N.~Patil, D.~Patterson, G.~Agrawal, R.~Bajwa, S.~Bates, S.~Bhatia, N.~Boden, A.~Borchers \emph{et~al.}, ``In-datacenter performance analysis of a tensor processing unit,'' in \emph{44th Annual International Symposium on Computer Architecture}, 2017, p. 1–12.

\bibitem{Eyeriss}
Y.-H. Chen, T.~Krishna, J.~Emer, and V.~Sze, ``Eyeriss: An energy-efficient reconfigurable accelerator for deep convolutional neural networks,'' in \emph{IEEE International Solid-State Circuits Conference, ISSCC 2016, Digest of Technical Papers}, 2016, pp. 262--263.

\bibitem{Gemmini}
H.~Genc, S.~Kim, A.~Amid, A.~Haj-Ali, V.~Iyer, P.~Prakash, J.~Zhao, D.~Grubb, H.~Liew, H.~Mao \emph{et~al.}, ``Gemmini: Enabling systematic deep-learning architecture evaluation via full-stack integration,'' in \emph{58th Annual Design Automation Conference (DAC)}, 2021.

\bibitem{Ahmadilivani2024}
M.~H. Ahmadilivani, M.~Taheri, J.~Raik, M.~Daneshtalab, and M.~Jenihhin, ``A systematic literature review on hardware reliability assessment methods for deep neural networks,'' \emph{ACM Comput. Surv.}, vol.~56, no.~6, 2024.

\bibitem{Iturbe2018}
X.~Iturbe, B.~Venu, E.~Ozer, J.-L. Poupat, G.~Gimenez, and H.-U. Zurek, ``The arm triple core lock-step {(TCLS)} processor,'' \emph{ACM Transactions on Computer Systems}, vol.~36, pp. 1--30, 2018.

\bibitem{Rogenmoser2022}
M.~Rogenmoser, N.~Wistoff, P.~Vogel, F.~Gurkaynak, and L.~Benini, ``On-demand redundancy grouping: Selectable soft-error tolerance for a multicore cluster,'' in \emph{IEEE Computer Society Annual Symposium on VLSI (ISVLSI)}, 2022, pp. 398--401.

\bibitem{Libano2019}
F.~Libano, B.~Wilson, J.~Anderson, M.~J. Wirthlin, C.~Cazzaniga, C.~Frost, and P.~Rech, ``Selective hardening for neural networks in {FPGAs},'' \emph{IEEE Transactions on Nuclear Science}, vol.~66, pp. 216--222, 1 2019.

\bibitem{Bolchini2022}
C.~Bolchini, L.~Cassano, A.~Miele, and A.~Nazzari, ``Selective hardening of {CNNs} based on layer vulnerability estimation,'' in \emph{IEEE International Symposium on Defect and Fault Tolerance in VLSI and Nanotechnology Systems (DFT)}, 2022.

\bibitem{Bertoa2023}
T.~G. Bertoa, G.~Gambardella, N.~J. Fraser, M.~Blott, and J.~McAllister, ``Fault-tolerant neural network accelerators with selective {TMR},'' \emph{IEEE Design and Test}, vol.~40, pp. 67--74, 4 2023.

\bibitem{Khoshavi2020}
N.~Khoshavi, A.~Roohi, C.~Broyles, S.~Sargolzaei, Y.~Bi, and D.~Z. Pan, ``{SHIELDeNN}: Online accelerated framework for fault-tolerant deep neural network architectures,'' in \emph{Design Automation Conference}, 7 2020.

\bibitem{Ahmadilivani2023}
M.~H. Ahmadilivani, M.~Taheri, J.~Raik, M.~Daneshtalab, and M.~Jenihhin, ``Enhancing fault resilience of {QNNs} by selective neuron splitting,'' in \emph{IEEE 5th International Conference on Artificial Intelligence Circuits and Systems (AICAS)}, 2023, pp. 1--5.

\bibitem{Zahid2020}
U.~Zahid, G.~Gambardella, N.~J. Fraser, M.~Blott, and K.~Vissers, ``{FAT}: Training neural networks for reliable inference under hardware faults,'' in \emph{International Test Conference}, 11 2020.

\bibitem{Jaicnao2015}
J.~Deng, Y.~Fang, Z.~Du, Y.~Wang, H.~Li, O.~Temam, P.~Ienne, D.~Novo, X.~Li, Y.~Chen, and C.~Wu, ``Retraining-based timing error mitigation for hardware neural networks,'' in \emph{Design, Automation and Test in Europe Conference (DATE)}, 2015, pp. 593--596.

\bibitem{Xu2019}
D.~Xu, K.~Xing, C.~Liu, Y.~Wang, Y.~Dai, L.~Cheng, H.~Li, and L.~Zhang, ``Resilient neural network training for accelerators with computing errors,'' in \emph{IEEE 30th International Conference on Application-specific Systems, Architectures and Processors (ASAP)}, 2019, pp. 99--102.

\bibitem{Ozen2020a}
E.~Ozen and A.~Orailoglu, ``Just say zero: Containing critical bit-error propagation in deep neural networks with anomalous feature suppression,'' in \emph{IEEE/ACM International Conference on Computer-Aided Design, Digest of Technical Papers, ICCAD}, 11 2020.

\bibitem{Ozen2020b}
E.~Ozen and A.~Orailoglu, ``Boosting bit-error resilience of {DNN} accelerators through median feature selection,'' \emph{IEEE Transactions on Computer-Aided Design of Integrated Circuits and Systems}, vol.~39, pp. 3250--3262, 11 2020.

\bibitem{Hoang2020}
L.-H. Hoang, M.~A. Hanif, and M.~Shafique, ``{FT-ClipAct}: Resilience analysis of deep neural networks and improving their fault tolerance using clipped activation,'' in \emph{Design, Automation and Test in Europe}, 2020.

\bibitem{Chen2021}
Z.~Chen, G.~Li, and K.~Pattabiraman, ``A low-cost fault corrector for deep neural networks through range restriction,'' in \emph{51st Annual IEEE/IFIP International Conference on Dependable Systems and Networks (DSN)}, 2021, pp. 1--13.

\bibitem{Geissler2023}
F.~Geissler, S.~Qutub, M.~Paulitsch, and K.~Pattabiraman, ``A low-cost strategic monitoring approach for scalable and interpretable error detection in deep neural networks,'' in \emph{Computer Safety, Reliability, and Security: 42nd International Conference, SAFECOMP}, 2023, p. 75–88.

\bibitem{Goldstein2021}
B.~F. Goldstein, V.~C. Ferreira, S.~Srinivasan, D.~Das, A.~S. Nery, S.~Kundu, and F.~M. Franca, ``A lightweight error-resiliency mechanism for deep neural networks,'' in \emph{International Symposium on Quality Electronic Design (ISQED)}, 4 2021, pp. 311--316.

\bibitem{Kung1982}
H.~T. Kung, ``Why systolic architectures?'' \emph{Computer}, vol.~15, no.~1, pp. 37--46, 1982.

\bibitem{Vacca2023}
E.~Vacca, S.~Azimi, and L.~Sterpone, ``A comprehensive analysis of transient errors on systolic arrays,'' in \emph{26th International Symposium on Design and Diagnostics of Electronic Circuits and Systems (DDECS)}, 2023, pp. 175--180.

\bibitem{Tan2023}
J.~Tan, Q.~Wang, K.~Yan, X.~Wei, and X.~Fu, ``{Saca-FI}: A microarchitecture-level fault injection framework for reliability analysis of systolic array based {CNN} accelerator,'' \emph{Future Generation Computer Systems}, vol. 147, pp. 251--264, 10 2023.

\bibitem{Samajdar2020}
A.~Samajdar, J.~M. Joseph, Y.~Zhu, P.~Whatmough, M.~Mattina, and T.~Krishna, ``A systematic methodology for characterizing scalability of {DNN} accelerators using {SCALE-Sim},'' in \emph{IEEE International Symposium on Performance Analysis of Systems and Software (ISPASS)}, 8 2020, pp. 58--68.

\bibitem{Agarwal2023}
U.~K. Agarwal, A.~Chan, A.~Asgari, and K.~Pattabiraman, ``Towards reliability assessment of systolic arrays against stuck-at faults,'' in \emph{53rd Annual IEEE/IFIP International Conference on Dependable Systems and Networks - Supplemental Volume, DSN-S}, 2023, pp. 230--236.

\bibitem{Taheri2024}
M.~Taheri, M.~Daneshtalab, J.~Raik, M.~Jenihhin, S.~Pappalardo, P.~Jimenez, B.~Deveautour, and A.~Bosio, ``{SAFFIRA}: a framework for assessing the reliability of systolic-array-based {DNN} accelerators,'' in \emph{27th International Symposium on Design and Diagnostics of Electronic Circuits and Systems (DDECS)}, 2024, pp. 19--24.

\bibitem{Zhao2022}
Y.~Zhao, K.~Wang, and A.~Louri, ``{FSA}: An efficient fault-tolerant systolic array-based {DNN} accelerator architecture,'' in \emph{IEEE International Conference on Computer Design: VLSI in Computers and Processors}, 2022, pp. 545--552.

\bibitem{Lee2024b}
H.~Lee, J.~Park, and S.~Kang, ``An area-efficient systolic array redundancy architecture for reliable ai accelerator,'' \emph{IEEE Transactions on Very Large Scale Integration (VLSI) Systems}, vol.~32, pp. 1950--1954, 10 2024.

\bibitem{Lee2023}
H.~Lee, J.~Kim, J.~Park, and S.~Kang, ``Strait: Self-test and self-recovery for ai accelerator,'' \emph{IEEE Transactions on Computer-Aided Design of Integrated Circuits and Systems}, vol.~42, pp. 3092--3104, 9 2023.

\bibitem{Zhang2019}
J.~J. Zhang, K.~Basu, and S.~Garg, ``Fault-tolerant systolic array based accelerators for deep neural network execution,'' \emph{IEEE Design and Test}, vol.~36, pp. 44--53, 10 2019.

\bibitem{Libano2023}
F.~Libano, P.~Rech, and J.~Brunhaver, ``Efficient error detection for matrix multiplication with systolic arrays on {FPGAs},'' \emph{IEEE Transactions on Computers}, vol.~72, pp. 2390--2403, 8 2023.

\bibitem{Vacca2023a}
E.~Vacca, G.~Ajmone, and L.~Sterpone, ``{RunSAFER}: A novel runtime fault detection approach for systolic array accelerators,'' in \emph{The 41st IEEE International Conference on Computer Design}, 2023.

\bibitem{Cora2025}
G.~Cora, E.~Vacca, C.~D. Sio, S.~Azimi, and L.~Sterpone, ``Repair: Reconfigurable platform for ai resilience within risc-v ecosystem,'' \emph{Lecture Notes in Computer Science}, vol. 15594 LNCS, pp. 71--87, 2025.

\bibitem{Peltekis2024}
C.~Peltekis, D.~Filippas, and G.~Dimitrakopoulos, ``Error checking for sparse systolic tensor arrays,'' in \emph{2024 IEEE 6th International Conference on AI Circuits and Systems (AICAS)}, 04 2024, pp. 552--556.

\bibitem{Peltekis2025}
C.~Peltekis, C.~Nicopoulos, and G.~Dimitrakopoulos, ``Periodic online testing for sparse systolic tensor arrays,'' in \emph{2025 14th International Conference on Modern Circuits and Systems Technologies (MOCAST)}.\hskip 1em plus 0.5em minus 0.4em\relax IEEE, 6 2025, pp. 1--6.

\bibitem{Safarpour2022}
M.~Safarpour, R.~Inanlou, and O.~Silvén, ``Algorithm level error detection in low voltage systolic array,'' \emph{IEEE Transactions on Circuits and Systems II: Express Briefs}, vol.~69, pp. 569--573, 2 2022.

\bibitem{Ibtesam2022}
M.~Ibtesam, U.~S. Solangi, J.~Kim, M.~A. Ansari, and S.~Park, ``Highly efficient test architecture for low-power ai accelerators,'' \emph{IEEE Transactions on Computer-Aided Design of Integrated Circuits and Systems}, vol.~41, pp. 2728--2738, 8 2022.

\bibitem{Lee2025}
H.~Lee, J.~Lee, and S.~Kang, ``A robust test architecture for low-power ai accelerators,'' \emph{IEEE Transactions on Computer-Aided Design of Integrated Circuits and Systems}, vol.~44, pp. 1581--1594, 4 2025.

\bibitem{Yin2018}
S.~Yin, P.~Ouyang, S.~Tang, F.~Tu, X.~Li, S.~Zheng, T.~Lu, J.~Gu, L.~Liu, and S.~Wei, ``A high energy efficient reconfigurable hybrid neural network processor for deep learning applications,'' \emph{IEEE Journal of Solid-State Circuits}, vol.~53, pp. 968--982, 4 2018.

\bibitem{Lee2024}
C.-J. Lee and T.~T. Yeh, ``{ReSA:} reconfigurable systolic array for multiple tiny dnn tensors,'' \emph{ACM Transactions on Architecture and Code Optimization}, vol.~37, p.~24, 9 2024.

\bibitem{Peltekis2023}
C.~Peltekis, D.~Filippas, G.~Dimitrakopoulos, C.~Nicopoulos, and D.~Pnevmatikatos, ``{ArrayFlex}: A systolic array architecture with configurable transparent pipelining,'' in \emph{Design, Automation and Test in Europe Conference (DATE)}, 2023.

\bibitem{Zhao2023}
T.~Zhao, S.~Miao, S.~Lu, J.~Cao, J.~Qiu, X.~Shi, K.~Wang, and L.~He, ``Towards a reconfigurable systolic array with multi-level packing for transformers,'' in \emph{Architecture and System Support for Transformer Models (ASSYST@ ISCA 2023)}, 2023.

\bibitem{Chen2016}
Y.-H. Chen, J.~Emer, and V.~Sze, ``Eyeriss: A spatial architecture for energy-efficient dataflow for convolutional neural networks,'' in \emph{ACM/IEEE 43rd Annual International Symposium on Computer Architecture (ISCA)}, 2016, pp. 367--379.

\bibitem{Xu2021}
R.~Xu, S.~Ma, Y.~Wang, and Y.~Guo, ``{HeSA:} heterogeneous systolic array architecture for compact {CNNs} hardware accelerators,'' in \emph{Design, Automation and Test in Europe (DATE)}, 2 2021, pp. 657--662.

\bibitem{Deng2020}
L.~Deng, G.~Li, S.~Han, L.~Shi, and Y.~Xie, ``Model compression and hardware acceleration for neural networks: A comprehensive survey,'' \emph{Proceedings of the IEEE}, vol. 108, no.~4, pp. 485--532, 2020.

\bibitem{Burel2022}
S.~Burel, A.~Evans, and L.~Anghel, ``{MOZART+}: Masking outputs with zeros for improved architectural robustness and testing of dnn accelerators,'' \emph{IEEE Transactions on Device and Materials Reliability}, vol.~22, no.~2, pp. 120--128, 2022.

\bibitem{Chellapilla2006}
K.~Chellapilla, S.~Puri, and P.~Simard, ``High performance convolutional neural networks for document processing,'' in \emph{10th International Workshop on Frontiers in Handwriting Recognition}, 2006.

\bibitem{Bolchini2023}
C.~Bolchini, L.~Cassano, A.~Miele, and A.~Toschi, ``Fast and accurate error simulation for {CNNs} against soft errors,'' \emph{IEEE Transactions on Computers}, vol.~72, pp. 984--997, 4 2023.

\bibitem{Mukherjee2003}
S.~S. Mukherjee, C.~Weaver, J.~Emer, S.~K. Reinhardt, and T.~Austin, ``A systematic methodology to compute the architectural vulnerability factors for a high-performance microprocessor,'' \emph{Annual International Symposium on Microarchitecture, MICRO}, vol. 2003-January, pp. 29--40, 2003.

\bibitem{Leveugle2009}
R.~Leveugle, A.~Calvez, P.~Maistri, and P.~Vanhauwaert, ``Statistical fault injection: Quantified error and confidence,'' in \emph{IEEE Design, Automation and Test in Europe Conference and Exhibition (DATE)}, 2009, pp. 502--506.

\end{thebibliography}

\end{document}